\documentclass{emulateapj}

\usepackage{color}

\slugcomment{Received 2012 October 22; accepted 2012 November 26; published 2013 January 4}

\shorttitle{Wide-field Multiband Photometry of Globular Cluster Systems}
\shortauthors{Kim et al.}

\begin{document}

\title{WIDE-FIELD MULTIBAND PHOTOMETRY OF
GLOBULAR CLUSTER SYSTEMS IN THE FORNAX GALAXY CLUSTER}

\author{Hak-Sub Kim\altaffilmark{1}, Suk-Jin Yoon\altaffilmark{1}, Sangmo Tony Sohn\altaffilmark{2}, Sang Chul Kim\altaffilmark{3}, Eunhyeuk Kim\altaffilmark{4},\\
Chul Chung\altaffilmark{1}, Sang-Yoon Lee\altaffilmark{1}, and Young-Wook Lee\altaffilmark{1}}

\affil{
\altaffilmark{1}Department of Astronomy \& Center for Galaxy Evolution Research, Yonsei University, Seoul 120-749, Republic of Korea; 
sjyoon@galaxy.yonsei.ac.kr\\
\altaffilmark{2}Space Telescope Science Institute, 3700 San Martin Drive, Baltimore, MD 21218, USA\\
\altaffilmark{3}Korea Astronomy and Space Science Institute, Daejeon 305-348, Republic of Korea\\
\altaffilmark{4}Korea Aerospace Research Institute, Daejeon 305-806, Republic of Korea}

\begin{abstract}
We present wide-field multiband photometry of globular cluster (GC) systems 
in NGC 1399, NGC 1404, and NGC 1387 located at the central region of the Fornax galaxy cluster.
Observation was carried out through $U$, $B$, $V$, and $I$ bands, 
which marks one of the widest and deepest $U$-band studies on extragalactic GC systems. 
The present $U$-band photometry enables us 
to significantly reduce the contamination by a factor of two for faint sources ($V_{0}$\,$\sim$\,23.5). 
The main results based on some 2000 GC candidates
around NGC 1399, NGC 1404, and NGC 1387 are as follows: 
(1) the GC system in each galaxy exhibits bimodal color distributions in all colors examined, 
but the shape of color histograms varies systematically depending on colors; 
(2) NGC 1399 shows that the mean colors of both blue and red GCs become bluer with increasing galactocentric radius;
(3) NGC 1399 shows overabundance of GCs in the directions of NGC 1404 and NGC 1387,
indicating their ongoing interactions;
and (4) NGC 1399 also exhibits a $\sim$\,0\farcm5~offset between the center of the inner GC distribution and the galaxy's optical center, 
suggesting that NGC 1399 is not yet dynamically relaxed and may be undergoing merger events.
\end{abstract}


\section{INTRODUCTION}

The Fornax galaxy cluster is the second nearest cluster of galaxies after Virgo, 
with a distance modulus of $(m-M)=31.51$ \citep{Blakeslee09}
corresponding to a distance of 20.0 Mpc. 
It consists of two major components of the main Fornax cluster centered on NGC 1399 
and a subcluster dominated by NGC 1316 \citep{Drinkwater01}. 
There have been many photometric and spectroscopic studies of globular cluster (GC) systems 
in the central region of the Fornax cluster, 
and a fair number of them concentrated on the GC system associated with NGC 1399.

\citet{Bridges91} and \citet{Wagner91} presented the first CCD photometry for NGC 1399 GCs. 
Both studies presented an unusually high GC specific frequency of $S_{N}>15$ 
(i.e., number of GCs normalized to the galaxy luminosity; see \citealt{Harris81}). 
Although subsequent studies (\citealt{Ostrov98, Dirsch03}) derived a lower value ($S_{N}\sim5$), 
it is still higher than those of normal elliptical galaxies. 
\citet{Forbes97} suggested that NGC 1404 GCs may have been tidally stripped by NGC 1399, 
resulting in a high $S_{N}$ value of NGC 1399 and a lower $S_{N}$ value of NGC 1404. 
\citet{Kissler99} also proposed that the over abundance of NGC 1399 GCs can be explained 
by tidal stripping of GCs from neighboring galaxies and by the accretion of GCs in the gravitational potential of the Fornax cluster. 
\citet{Grillmair99} supported this idea by showing that GCs in NGC 1399 and NGC 1404 
are statistically indistinguishable in terms of their color distributions and luminosity functions in their $B, I$ photometry. 
\citet{Bekki03} demonstrated the tidal stripping and accretion of GCs between NGC 1399 and NGC 1404 
through their numerical simulations. 
\citet{Bassino06} presented the asymmetries in the azimuthal distribution of NGC 1399 GCs 
and suggested that they may be explained by the tidal stripping of GCs from NGC 1387.

Since \citet{Ostrov93} first suggested a possible bimodal $C-T1$ color distribution in the GC system of NGC 1399, 
many studies have confirmed the bimodal or multimodal GC color distribution in this galaxy 
through various passbands (e.g., \citealt{Ostrov98, Grillmair99}).
Several studies have investigated the properties of blue and red GCs separately.
\citet{Forbes98} and \citet{Ostrov98} showed that red GCs appear more centrally concentrated than blue GCs. 
\citet{Forte05} suggested that the distribution of blue GCs is similar to that inferred for dark matter.
\citet{Dirsch03} presented that the surface density profile of blue GCs is shallower than that of red GCs within 8$\arcmin$ 
but they are not distinguishable beyond the radius. 
They also showed that the specific frequency of blue GCs is a factor of three larger than that of the red GCs within 7$\arcmin$. 
\citet{Forte05} suggested that the specific frequencies for red and blue GCs are 3.3 and 14.3, respectively. 
\citet{Grillmair99} showed that the peak of luminosity function for blue GCs is 0.36 mag brighter in $B$ than that for red GCs. 
\citet{Blakeslee12} found that the optical--infrared color distribution of NGC 1399 GCs is unimodal,
whereas the optical colors of the GCs exhibit bimodality.
\citet{Forte07} investigated the connection between GCs and the stellar population of NGC 1399 using their $C-T1$ photometry.
They suggested that the metal-poor subpopulation is homogeneous and exhibits an extended spatial distribution 
contributing $\sim$\,20\% of the total stellar mass, while the metal-rich subpopulation is heterogeneous and 
dominates the inner region of the galaxy.

\begin{deluxetable}{ccccc}
\tablecolumns{5}
\tablecaption{Host Galaxy Properties} 
\tablewidth{0pt}
\tablehead{\colhead{Galaxy} & {R.A. (J2000)} & {Decl. (J2000)} & {$B_{T}$} & {Type}\\ 
{} & {(h:m:s)} & {(\degr:\arcmin:\arcsec)} & {(mag)} & {}}
\startdata
NGC 1387 & 03:36:56.84 & --35:30:23.85 & 12.3 & SB0 \\ 
NGC 1389 & 03:37:11.67 & --35:44:39.73 & 12.8 & SB0 \\ 
NGC 1399 & 03:38:29.14 & --35:27:02.30 & 10.6 & E0 \\
NGC 1404 & 03:38:52.08 & --35:35:37.67 & 10.9 & E2 \\
\enddata
\label{tbl-target}
\end{deluxetable}

Recent studies based on the Advanced Camera for Surveys (ACS) Fornax Cluster Survey (FCS; \citealt{Jordan07}) 
have looked into various properties of GC systems in the Fornax galaxy cluster.
\citet{Mieske10} detected color--magnitude relation for GCs (i.e., blue tilt) 
and found that the slope is shallower than that found for Virgo Cluster Survey GC samples.
\citet{Masters10} measured the half-light radii of GCs in ACS FCS galaxies, 
and suggested the mean half-light radius as a distance indicator.
\citet{Villegas10} found that the dispersion of the GC luminosity functions varies systematically with the brightness of parent galaxies.
\citet{Liu11} investigated the radial color gradients of GC systems in early-type galaxies in the Fornax galaxy cluster, 
and found that both red and blue GC subpopulations exhibit significant color gradients.
They also found that the slope of the GC color gradients depends on mass of host galaxies.

There have been a few studies for the GC systems of NGC 1404 and NGC 1387.
For NGC 1404, \citet{Richtler92} presented $V$, $R$ photometry of the GC system, 
and derived a low specific frequency of $S_{N}=2.3$. They suggested that 
NGC 1404 may have its origin outside the Fornax cluster 
based on the low specific frequency similar to that of isolated galaxies						
together with its high radial velocity.
\citet{Forbes98} presented broad $B-I$ distributions for GCs in the inner and outer regions of NGC 1404 
and tentatively suggested a presence of two GC subpopulations. 
The bimodal distribution in $B-I$ was later confirmed by \citet{Grillmair99} and \citet{Larsen01}.  						
For NGC 1387, \citet{Bassino06b} found that the $C-T1$ distribution of the GCs presents 
a distinct separation between red and blue GCs, and the radial distribution of red GCs
is more centrally concentrated than blue GCs.

\begin{deluxetable}{ccccc}
\tablecolumns{5}
\tablecaption{Observation Log}
\tablewidth{0pt}
\tablehead{\colhead{Date (UT)} & {Filter} & {Exposure Time} & {Number of} & {Seeing} \\
{} & {} & {(s)} & {Frames} & {(arcsec)}}
\startdata
2006 Nov 24 & $U$\tablenotemark{a} & 1800 & 6 & 0.81-1.08 \\
2006 Nov 24 & $B$ & 900 & 3 & 0.91--1.20 \\
2006 Nov 24 & $V$ & 240 & 3 & 0.94--1.10 \\
2006 Nov 24 & $I$ & 300 & 6 & 0.63--0.80 \\
2006 Nov 28 & $U$\tablenotemark{a} & 1800 & 9 & 1.14--1.43 \\
2006 Nov 28 & $B$ & 900 & 3 & 1.15--1.35 \\
2006 Nov 28 & $V$ & 360 & 3 & 0.86--1.07 \\
\enddata
\tablenotetext{a}{We used the SDSS $u$ filter and later calibrated the photometry to the Johnson $U$ system.}
\label{tbl-obs}
\end{deluxetable}

Several studies have been done on kinematics and dynamics for GCs around NGC 1399. 
\citet{Grillmair94} suggested that, based on the low-dispersion spectra of 47 GCs, 
the large velocity dispersion of the GCs indicates 
that they are associated with the gravitational potential of the Fornax cluster rather than with that of NGC 1399. 
\citet{Kissler99} supported this suggestion by finding a strong radial increase of the velocity dispersion for 74 GCs. 
They found no compelling evidence for rotation and could not detect differences between the kinematics of the blue and red GCs. 
\citet{Richtler04} obtained medium-resolution spectra of 468 GCs, and they did not find any signature of rotation. 
They suggested that the large velocity dispersion of the GCs could arise from the contamination by foreground objects.
\citet{Schuberth10} investigated kinematical and dynamical properties  
of the GC system of NGC 1399 using the spectra of some 700 GCs. 
They found that blue and red populations are kinematically distinct, 
and the properties of red population resemble those of field stellar populations.

Several spectroscopic studies investigated chemical abundance and ages of NGC 1399 GCs.   
\citet{Kissler98} obtained spectra of 18 GCs in NGC 1399 
and found that their abundances span the range observed in the Milky Way and M31. 
They pointed out that $V-I$ correlates well with metallicity but the slope of the relation
becomes flatter toward redder colors if the values of Milky Way GCs are considered together.
\citet{Forbes01} investigated the ages of the GCs and found 
that the majority of their sample (10 GCs) have ages of around 11 Gyr.

In this paper we provide a new $U$, $B$, $V$, and $I$ photometric catalog of GC candidates 
in the central region of the Fornax galaxy cluster.
This is one of the widest and deepest $U$-band studies on extragalactic GC systems. 
Section 2 describes the observations and data reduction.
Section 3 gives the GC catalog for NGC 1399, NGC 1404, and NGC 1387, 
discussing the selection method and the foreground and background contamination level.
Section 4 presents the properties of GC systems around the galaxies, 
including spatial distributions, color--magnitude diagrams (CMDs), color distributions, 
and the radial variations of color bimodality.
Section 5 summarizes our results.

\section{OBSERVATIONS AND DATA REDUCTION}
\subsection{Observations}

The observations were carried out on the nights of 2006 November 24 
and 28 (UT) with the Mosaic II CCD imager mounted on the prime focus of the 
4 m Blanco telescope at Cerro Tololo Inter-American Observatory (CTIO). 
The observing conditions were clear and photometric with an average 
seeing of around 1$\arcsec$. The Mosaic II consists of eight 
2048$\times$4096 pixel CCDs with a pixel scale of 0\farcs27 
providing a total field of view of 36$\arcmin\times$36$\arcmin$. 

\begin{figure*}
\epsscale{1.1}
\plotone{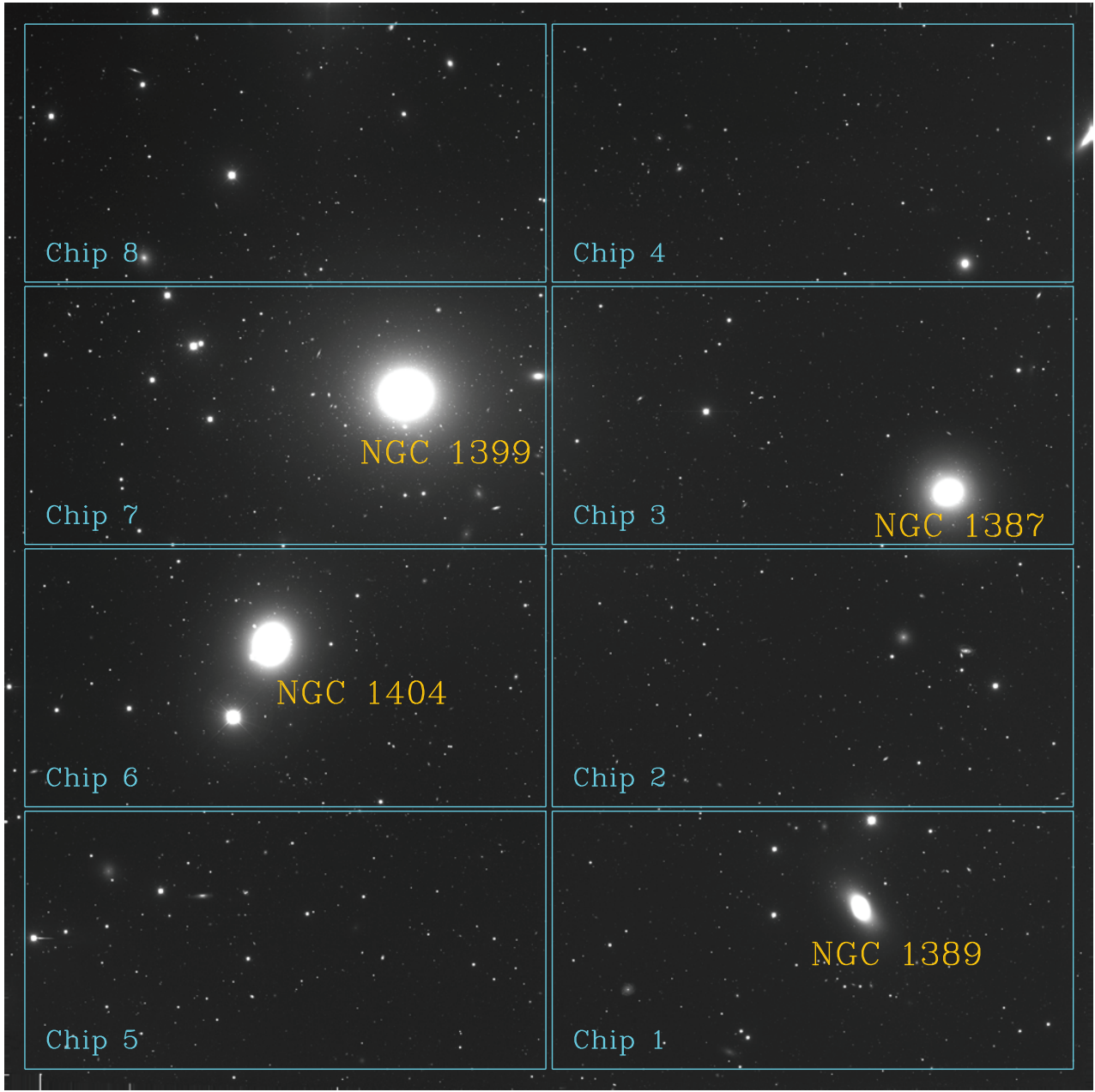}
\caption{Median combined $B$-band image of our target field 
with the configuration of Mosaic II chips overlaid. 
The gaps between chips were filled in by creating mosaics using a three-point dither pattern.
The image covers an area of $36\arcmin\times36\arcmin$, 
containing four Fornax cluster galaxies as labeled and their GC systems.  
North is up, and east is to the left.\\
\label{fig1_N1399B}}
\end{figure*}

Our target field was chosen such that the brightest elliptical galaxy 
in the Fornax cluster, NGC 1399, and its three neighboring galaxies, 
NGC 1404, NGC 1387, and NGC 1389,
were all imaged in a single telescope 
pointing. Figure~\ref{fig1_N1399B} shows the median combined image of our target 
field as imaged in the $B$-filter, and Table~\ref{tbl-target} lists the basic 
properties of the four members of the Fornax cluster. We observed the 
target field with $u$, $B$, $V$, and $I$ filters. We note that the 
standard Johnson $U$ filter was not available at the time of our 
observations, so we used the Sloan Digital Sky Survey (SDSS) $u$ filter as an alternative and
later calibrated the photometry to the Johnson $U$ system. The 
resulting effect of this calibration procedure is discussed below. 
The total exposure times in $u$, $B$, $V$, and $I$ were 27000, 5400, 
1800, and 1800 s, respectively. We split our observations in each 
band into two sets of three dithered exposures (five sets for $u$) to 
fill in the gaps between the individual CCD chips in the mosaic, 
and to minimize the impact on our photometry from CCD blemishes 
(e.g., hot pixels and bad columns). A summary of observations is 
provided in Table~\ref{tbl-obs}.

\subsection{Pre-processing and Photometry}

\begin{figure*}
\epsscale{1.0}
\plotone{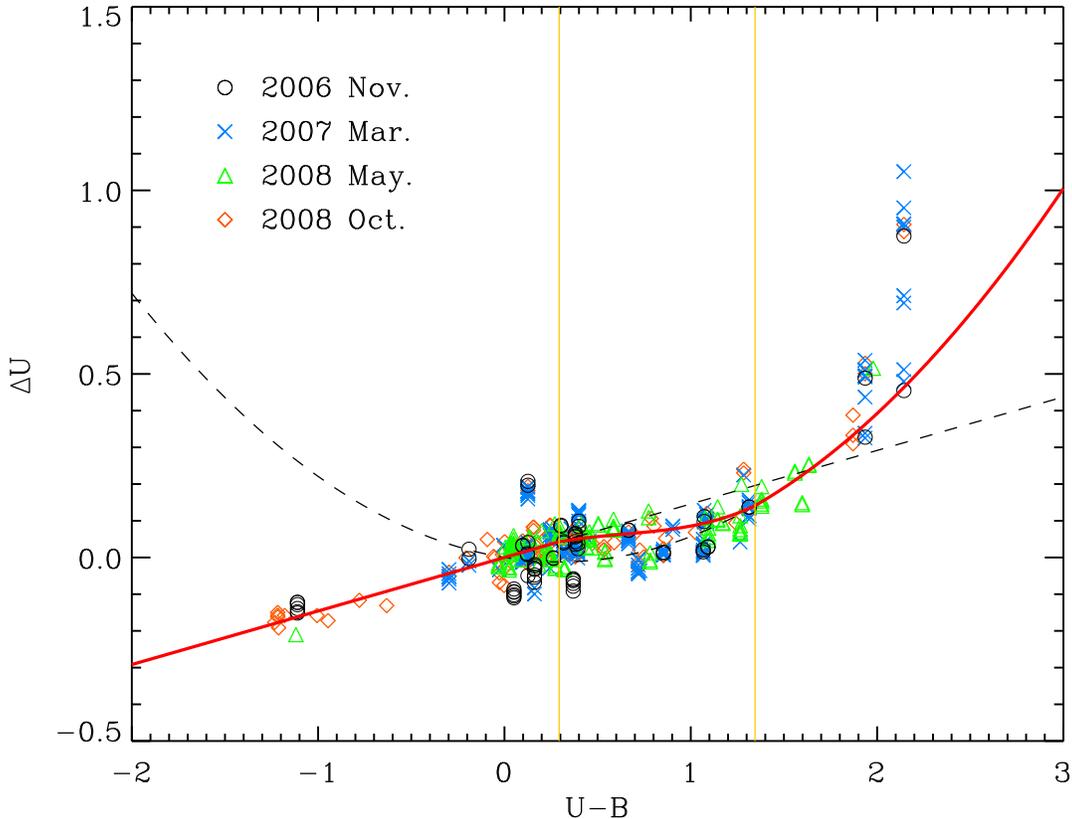}
\caption{
Difference between the standard and instrumental $U$ magnitudes of standard stars
after photometric zero point and atmospheric extinction corrections,
as a function of $U-B$. 
In the calibration of $U$ band, all the standard stars         
observed in subsequent observing runs during three years
were used to derive a reliable solution.
Different symbols are used to distinguish observations from different seasons.
Two dashed lines represent error-weighted least-squares fits for blue and red standard stars, respectively.
Red solid line represents the adopted best-fit function, 
which connects the two functions  
using internal division points of them between two vertical orange lines. \\
}
\label{fig2-STD}
\end{figure*}

All images were pre-processed using the MSCRED package 
\citep{Valdes98} in IRAF.\footnote{\footnotesize IRAF is distributed by the National 
Optical Astronomy Observatory, which is operated by the Association of 
Universities for Research in Astronomy (AURA) under cooperative 
agreement with the National Science Foundation.} We used the CCDPROC
task to correct for cross-talk between the CCD chips, apply and trim 
the overscan, correct for the bias level, and apply the flat-field 
correction. We used MSCPIXAREA to correct for the variations in pixel 
scales across the frame produced by geometric distortion in the 
Mosaic II imager. Each mosaic image was then split into eight 
individual images using the MSCSPLIT task.

To produce a clean high signal-to-noise image in each filter, we 
combined the images using the MONTAGE2 routine in the ALLFRAME package 
\citep{Stet94}. We then applied a ring median filter \citep{Secker95}
having an inner radius of 2.5 times the mean full width at half-maximum (FWHM)   
and a width of 5 pixel
on the combined images and subtracted off these images from the 
original images to remove the rapidly varying background light of the 
galaxies while preserving the point-sources. Objects were detected 
from the background-subtracted images using the DAOPHOT II/ALLSTAR 
routines through the following steps. First, we derived a list of 
objects and created new images with these objects subtracted. From 
these images, we detected fainter objects that were missed in the 
previous step and created another set of object-subtracted images.
This procedure was iterated five times, and all detected objects were 
merged into a master catalog. The master catalog was then input to 
the ALLFRAME program \citep{Stet94} along with all of the images 
and their point-spread function (PSF) models for final PSF photometry.
Aperture corrections were computed through the DAOGROW routine 
\citep{Stet90} using bright and isolated point-sources, and we 
applied them to our photometry above. Finally, an error-weighted 
mean instrumental magnitude for each object was obtained using the 
DAOMATCH/DAOMASTER routines \citep{Stet93}.

\begin{deluxetable*}{cccccc}
\tablecolumns{6}
\tablecaption{Photometric Calibration Coefficients} 
\tablewidth{0pt}
\tablehead{\colhead{Date (UT)} & {Filter} & {c1} & {c2} & {c3} & {c4}}
\startdata
2006 Nov 24 & $U_{\textnormal{\tiny{blue}}}$\tablenotemark{a} &  --2.1194 $\pm$ 0.0127 & --0.4495 $\pm$ 0.0072 & 0.1458 $\pm$ 0.0038 & NaN \\
&$U_{\textnormal{\tiny{red}}}$\tablenotemark{a} & --2.0699 $\pm$ 0.0380 & --0.4442 $\pm$ 0.0144 & --0.0819 $\pm$ 0.0581 & 0.1390 $\pm$ 0.0259 \\
&$B$ & 0.3711 $\pm$ 0.0045 & --0.2236 $\pm$ 0.0037 & 0.0803 $\pm$ 0.0015 & NaN \\
&$V$ & 0.6402 $\pm$ 0.0035 & --0.1386 $\pm$ 0.0028 & --0.0297 $\pm$ 0.0011 & NaN \\
&$I$ & 0.0578 $\pm$ 0.0039 & --0.0576 $\pm$ 0.0030 & --0.0099 $\pm$ 0.0011 & NaN \\
\\
2006 Nov 28 & $U_{\textnormal{\tiny{blue}}}$\tablenotemark{a} & --2.1641 $\pm$ 0.0128 & --0.4495 $\pm$ 0.0072 & 0.1458 $\pm$ 0.0038 & NaN \\
&$U_{\textnormal{\tiny{red}}}$\tablenotemark{a} & --2.1196 $\pm$ 0.0383 & --0.4442 $\pm$ 0.0144 & --0.0819 $\pm$ 0.0581 & 0.1390 $\pm$ 0.0259 \\
&$B$ & 0.3243 $\pm$ 0.0048 & --0.2236 $\pm$ 0.0037 & 0.0803 $\pm$ 0.0015 & NaN \\
&$V$ & 0.6061 $\pm$ 0.0037 & --0.1386 $\pm$ 0.0028 & --0.0297 $\pm$ 0.0011 & NaN \\
&$I$ & 0.0350 $\pm$ 0.0041 & --0.0576 $\pm$ 0.0030 & --0.0099 $\pm$ 0.0011 & NaN \\
\enddata
\tablenotetext{a}{For $U$ band, two functions are used to find the best-fit (see the text).\\}
\label{tbl-STD}
\end{deluxetable*}

\begin{deluxetable}{cccc}
\tablecolumns{4}
\tablecaption{Chip-to-chip Correction Coefficients} 
\tablewidth{0pt}
\tablehead{\colhead{Chip} & {Filter} & {c1} & {c2}}
\startdata
Chip 1 & $U$ &  0.0347 $\pm$ 0.0020  & --0.0334 $\pm$ 0.0055  \\
       & $B$ &  0.0236 $\pm$ 0.0007  & --0.0147 $\pm$ 0.0022  \\
       & $V$ & --0.0165 $\pm$ 0.0004  &  0.0029 $\pm$ 0.0014  \\
       & $I$ &  0.0037 $\pm$ 0.0005  &  0.0010 $\pm$ 0.0017  \\
\\
Chip 2 & $U$ &  0.0323 $\pm$ 0.0019  & --0.0609 $\pm$ 0.0055  \\
       & $B$ &  0.0162 $\pm$ 0.0007  & --0.0094 $\pm$ 0.0024  \\
       & $V$ &  0.0034 $\pm$ 0.0005  &  0.0046 $\pm$ 0.0015  \\
       & $I$ &  0.0023 $\pm$ 0.0005  & --0.0037 $\pm$ 0.0017  \\
\\
Chip 3 & $U$ &  0.0087 $\pm$ 0.0019  & --0.0675 $\pm$ 0.0057  \\
       & $B$ &  0.0197 $\pm$ 0.0007  & --0.0125 $\pm$ 0.0021  \\
       & $V$ &  0.0020 $\pm$ 0.0004  &  0.0038 $\pm$ 0.0014  \\
       & $I$ &  0.0085 $\pm$ 0.0005  &  0.0013 $\pm$ 0.0016  \\
\\
Chip 4 & $U$ &  0.0208 $\pm$ 0.0020  & --0.0156 $\pm$ 0.0062  \\
       & $B$ &  0.0195 $\pm$ 0.0007  & --0.0093 $\pm$ 0.0021  \\
       & $V$ & --0.0060 $\pm$ 0.0005  &  0.0016 $\pm$ 0.0015  \\
       & $I$ & --0.0055 $\pm$ 0.0005  &  0.0026 $\pm$ 0.0016  \\
\\
Chip 5 & $U$ &  0.0409 $\pm$ 0.0020  &  0.0013 $\pm$ 0.0057  \\
       & $B$ & --0.0024 $\pm$ 0.0007  & --0.0018 $\pm$ 0.0022  \\
       & $V$ & --0.0149 $\pm$ 0.0005  &  0.0022 $\pm$ 0.0016  \\
       & $I$ &  0.0019 $\pm$ 0.0007  & --0.0006 $\pm$ 0.0014  \\
\\
Chip 7 & $U$ & --0.0183 $\pm$ 0.0020  & --0.0016 $\pm$ 0.0060  \\
       & $B$ & --0.0052 $\pm$ 0.0007  &  0.0047 $\pm$ 0.0021  \\
       & $V$ & --0.0023 $\pm$ 0.0004  & --0.0002 $\pm$ 0.0014  \\
       & $I$ &  0.0031 $\pm$ 0.0005  &  0.0033 $\pm$ 0.0017  \\
\\
Chip 8 & $U$ &  0.0136 $\pm$ 0.0020  & --0.0069 $\pm$ 0.0060  \\
       & $B$ &  0.0083 $\pm$ 0.0007  & --0.0037 $\pm$ 0.0021  \\
       & $V$ & --0.0046 $\pm$ 0.0005  &  0.0005 $\pm$ 0.0015  \\
       & $I$ &  0.0122 $\pm$ 0.0005  &  0.0035 $\pm$ 0.0017  \\
\enddata
\label{tbl-c2c}
\end{deluxetable}

The photometric calibration was achieved using standard stars in 
the fields of \citet{Lan92} and \citet{Stet00}. During the course of 
our observing runs, we imaged eight standard fields placing the stars 
of interest on chip 6. Each image contains 15--25 standard stars, 
which covers an appropriately large range in color. We also observed 
the standard fields in many different airmasses. 
In the calibration of $U$ band, however, we used all the standard stars 
that are observed in subsequent observing runs during 2006, 2007, and 2008, 
because the photometric uncertainty in $U$ band is quite large and 
the number of standard stars is too small to derive reliable solutions.
To calibrate the instrumental magnitudes to the Johnson-Cousins standard magnitudes,  
we used the transformation equation of the form 
\begin{equation} \label{eq:std}
M_{\textnormal{\tiny{std}}}=m_{\textnormal{\tiny{inst}}}+\textnormal{c1}+\textnormal{c2}\times \textnormal{airmass}+\textnormal{c3} \times \textnormal{color}+\textnormal{c4} \times \textnormal{color}^2,
\end{equation}
where ${M}_{\textnormal{\tiny{std}}}$ is the calibrated magnitude, 
$m_{inst}$ is the instrumental magnitude that is normalized to 1 s,
c1 is the photometric zero point offset,
c2 is the extinction coefficient for airmass corrections,
and both c3 and c4 are the color coefficients. 
The colors used in the calibration 
are $U-B$, $B-V$, $B-V$, and $V-I$ for $U$, $B$, $V$, and $I$, respectively.
The photometric coefficients are determined by an iterative process of   
error-weighted fitting of the transformation equation to the standard magnitudes.
The quadratic color term is included only in the equation of $U$ band, 
because a significant deviation from a linear fit
occurs in the transformation of $U$ band.
Figure~\ref{fig2-STD} shows the $U$ magnitude difference between 
the standard and instrumental magnitudes corrected for both of the photometric zero point and 
the atmospheric extinction, as a function of $U-B$. 
The dashed lines represent two best-fit functions for blue and red standard stars, respectively.   
However, if we divide the observed data into two groups by $U-B$ and use the two functions in calibration separately,
the discontinuity between two functions may lead to an artificial clump in color distributions of the data.
To avoid this problem, we connected the two functions (red lines)
at the internal division points of them                
in the range $0.29 < U-B < 1.35$ (two vertical lines). 
Finally, the instrumental magnitudes were converted to the standard Johnson-Cousins $UBVI$
system using the set of photometric coefficients listed in Table~\ref{tbl-STD}

\begin{figure*}
\epsscale{1.0}
\plotone{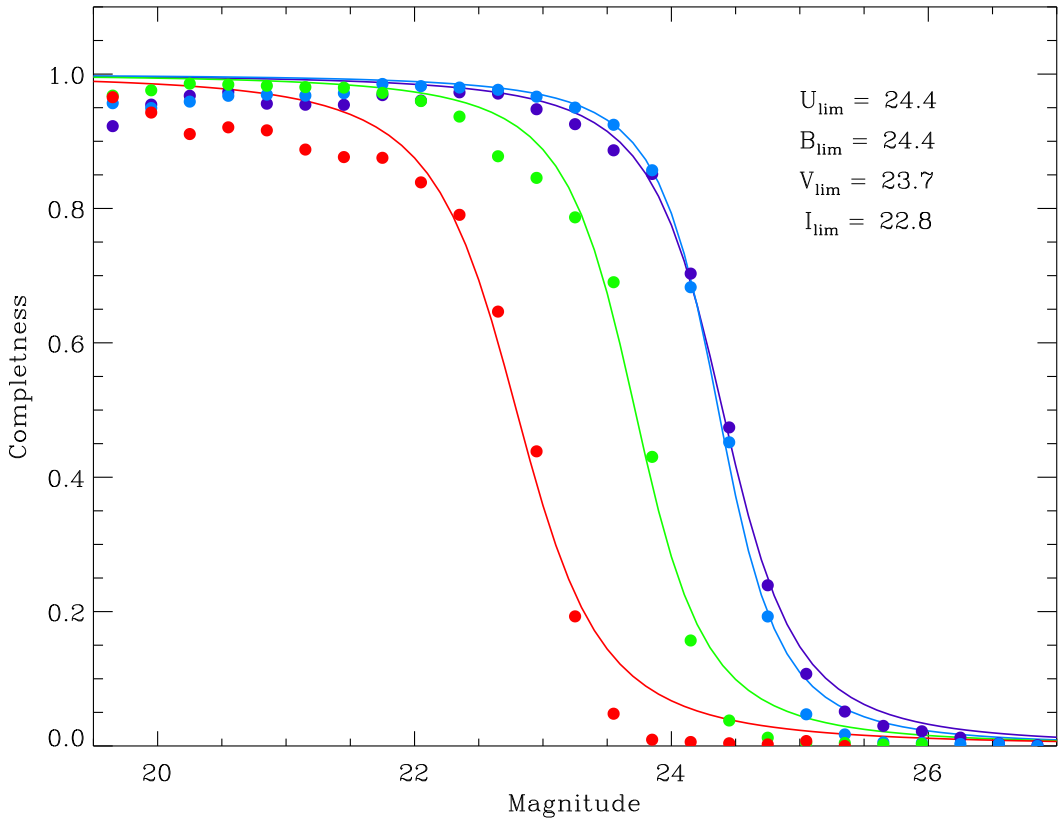}
\caption{Completeness functions obtained from artificial star tests. 
Detection efficiencies (filled circle) were fitted to the analytic function by \citet{Fleming95}, 
which are denoted by solid lines.
The purple, blue, green, and red lines represent the $U$, $B$, $V$, and $I$ bands, respectively. 
The functions are for the overall image, 
except the galactic central regions and the gaps between mosaic chips 
where the detection efficiency rapidly drops. 
Limiting magnitudes where the completeness drops to 50\,\% are indicated in the upper-right corner.\\
\label{fig3_Comtest}}
\end{figure*}

Calibrating the SDSS $u$ photometry to the Johnson $U$ system may cause
systematic shifts in the colors based on this band since the 
system throughputs are different between the two filters. 
To estimate these shifts, we simulated our observations
using the system throughputs of the two filters, synthetic spectra
of GCs from the YEPS model \citep{Chung12}, and stars from the Kurucz ATLAS9 model.
We calculated the magnitude through each filter for each model GC, 
and converted it to the standardized Johnson $U$ magnitude 
using the synthetic spectrum of a star having the same color of the GC.
We then calculated the differences between these standardized magnitudes 
from the respective filters. The differences range from 0.002 mag to 0.022 mag, 
depending on the colors of the model GCs. We conclude that these are small 
enough to have negligible impact  on our photometric analyses throughout the paper.

As the individual CCD chips of the Mosaic may have different 
color terms, we attempted to correct for this ``chip-to-chip variation'' 
as follows. On the night of 2008 May 5, we took multiple $uBVI$ 
observations of the Landolt standard fields SA98 and SA107 with the 
Mosaic II, such that the same set of standard stars was positioned on 
each chip. Using these data, we derived the color terms that correct 
for the chip-to-chip variation. Although these color terms were very 
small, we applied them to our data in hopes of making the multiple 
measurements of the same source more consistent.
The equation used in correction is of the form
\begin{equation} \label{eq:c2c}
M_{\textnormal{\tiny{cor}}}=m+\textnormal{c1}+\textnormal{c2} \times \textnormal{color}, 
\end{equation}
where $M_{\textnormal{\tiny{cor}}}$ is the magnitude converted to that on the reference chip 6, 
$m$ is the magnitude measured on each chip,
c1 is the zero point offset,
and c2 is the color coefficient.
The colors used in the calibration 
are $U-B$, $B-V$, $B-V$, and $V-I$ for $U$, $B$, $V$, and $I$, respectively.
Table~\ref{tbl-c2c} lists the zero point offsets and color coefficients for each chip.

We corrected for the foreground Galactic extinction using the 
reddening maps provided by \citet{Schlegel98}. To account for the 
atmospheric dispersion corrector (ADC) of the 4 m Blanco telescope 
that significantly modifies the system throughputs, we computed the
effective wavelengths of our four bands from the total system 
throughputs including the effect of the ADC, and we derived 
$A_{\lambda}/E(B-V)$ values based on the extinction law provided 
by \citet{Cardelli89}. 
The mean extinction values in $U$, $B$, $V$, and $I$ are 0.06, 0.05, 0.04, and 0.02 mag, respectively.

Our images were calibrated for astrometry using the USNO-B 1.0 
catalog stars \citep{Monet03} and the MSCTPEAK task. The typical 
rms of our astrometric solutions is at the level of 0\farcs2.
The R.A./decl. coordinates of our subsequent source catalogs use the 
positions derived from these images.

\begin{figure*}
\epsscale{0.75}
\plotone{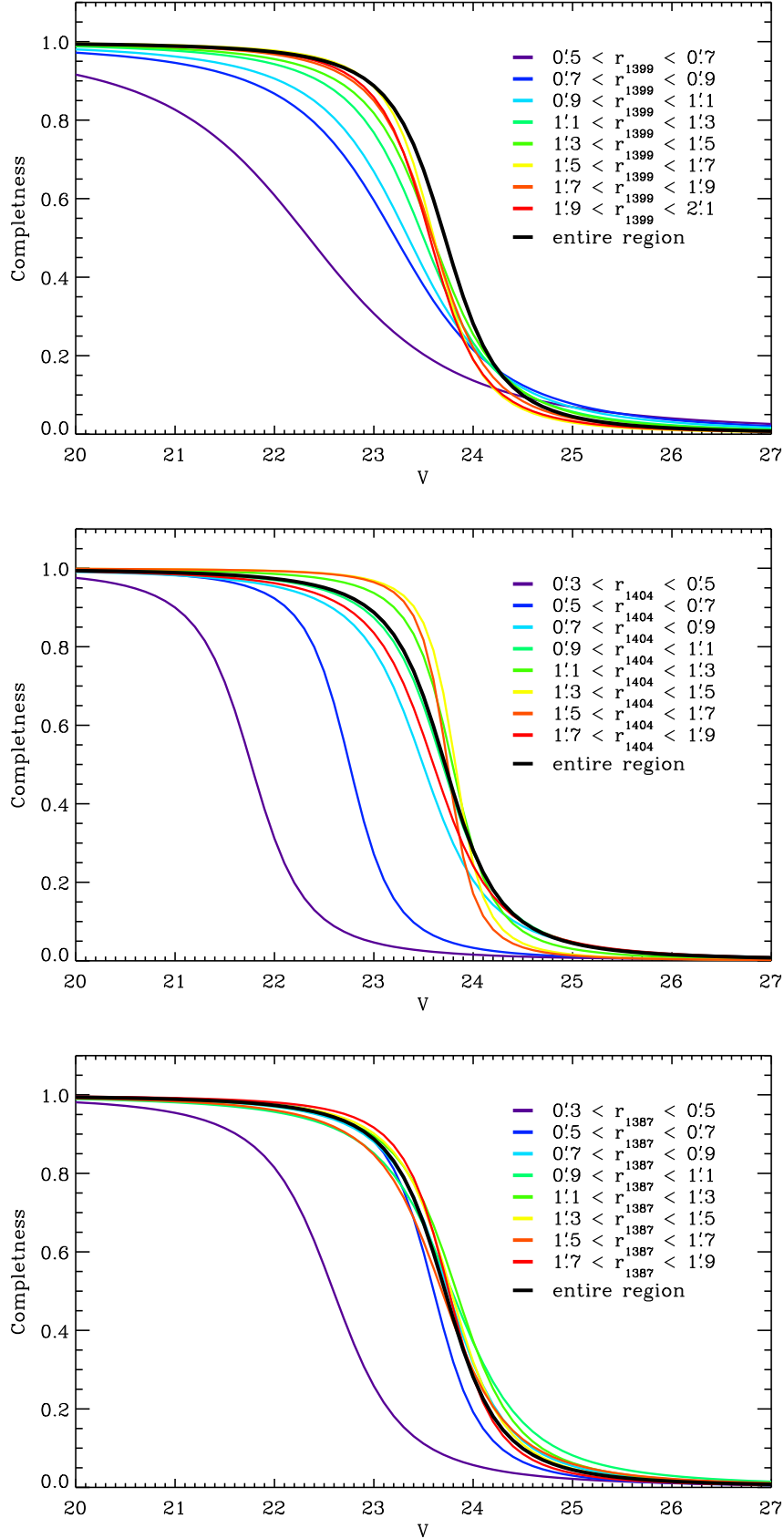}
\caption{ $V$-band completeness functions at various radial bins   
from the center of galaxies---NGC 1399 (top), NGC 1404 (middle), and NGC 1387 (bottom). 
The black thick line represents the overall completeness function. 
The completeness is very low near the center of the galaxies,
and increases as one moves out from the center,
and then flattens out beyond 1\farcm3 for NGC 1399, 
0\farcm7 for NGC 1404, and 0\farcm5 for NGC 1387.
\label{fig4_Com_radial}}
\end{figure*}

\begin{figure*}
\epsscale{1.0}
\plotone{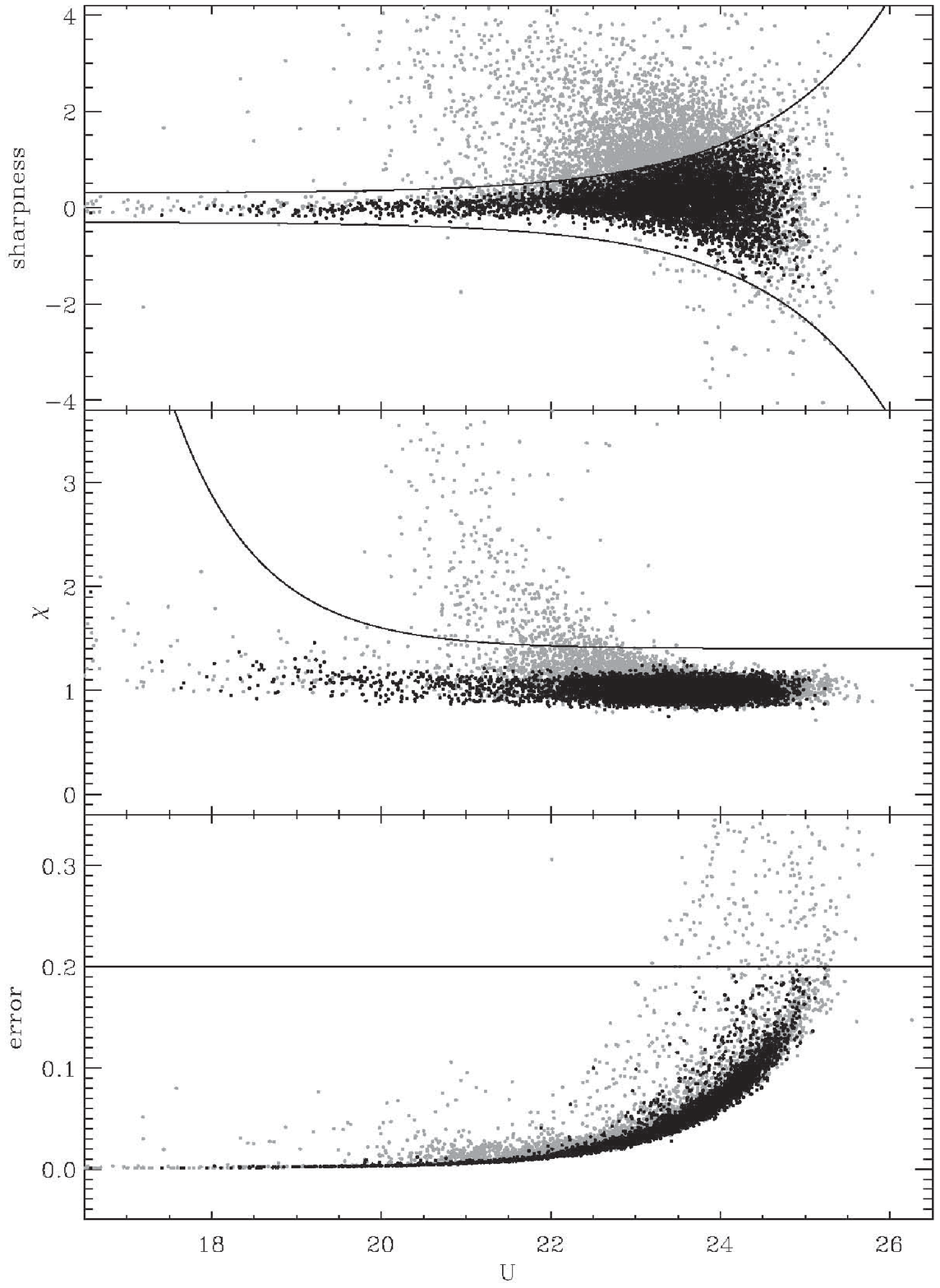}
\caption{Photometric error and image parameters ($\chi$ and $Sharpness$) 
in the $U$ band as a function of magnitude. 
Solid lines show the point-source selection criteria. 
Black dots represent the selected sources with good photometry, 
while gray dots represent the sources that do not satisfy any of the selection criteria. \\
\label{fig5_chi}}
\end{figure*}

\subsection{Completeness}

We performed artificial star tests to estimate the completeness of 
our photometry. We used the ADDSTAR routine of DAOPHOT II to generate 
4000 artificial stars in each chip
with magnitudes and color ranges 
consistent with those of the point-sources detected and measured above. 
The luminosity function of our artificial stars was chosen to 
be flat to preserve statistical significance in all magnitude ranges.
Artificial stars were added to the PSF-subtracted images using the 
empirical PSFs derived during our photometry. We used the same 
photometry procedure as above to detect and measure artificial 
stars in these images. This process was repeated seven times 
with the positions of artificial stars altering randomly each time,
resulting in a total of 28,000 artificial stars per chip.
In each 0.3 mag bin, we calculated the detection efficiency, 
i.e., the ratio of the number of recovered artificial stars to 
the number of added ones. Central regions of galaxies and the gaps 
between CCD chips were masked for the calculation. 
To obtain a limiting magnitude at which the completeness drops to 50\,\%, 
the detection efficiencies were fitted to
the analytic function by \citet{Fleming95},       
\begin{equation} \label{eq:compl}
f(m)=\frac{1}{2}\Big(1-\frac{\alpha(m-m_{\textnormal{\tiny{lim}}})}{\sqrt{1+\alpha^2(m-m_{\textnormal{\tiny{lim}}})^2}}\Big),
\end{equation} 
where $m_{\textnormal{\tiny{lim}}}$ is the limiting magnitude, 
and $\alpha$ is the parameter that controls 
the steepness of the completeness function near the limiting magnitude. 
Figure~\ref{fig3_Comtest} shows the measured completeness (dots) and the fitted 
functions (solid lines) in $U$, $B$, $V$, and $I$ bands depicted by 
purple, blue, green, and red colors, respectively. The limiting 
magnitudes of the four bands are 24.4, 24.4, 23.7, and 22.8 mag in     
$U$, $B$, $V$, and $I$ bands, respectively. We note that the limiting 
magnitudes slightly vary in the $<$\,0.2 mag level among 
individual CCDs due to variations in detector characteristics.

\begin{figure}
\epsscale{1.12}
\plotone{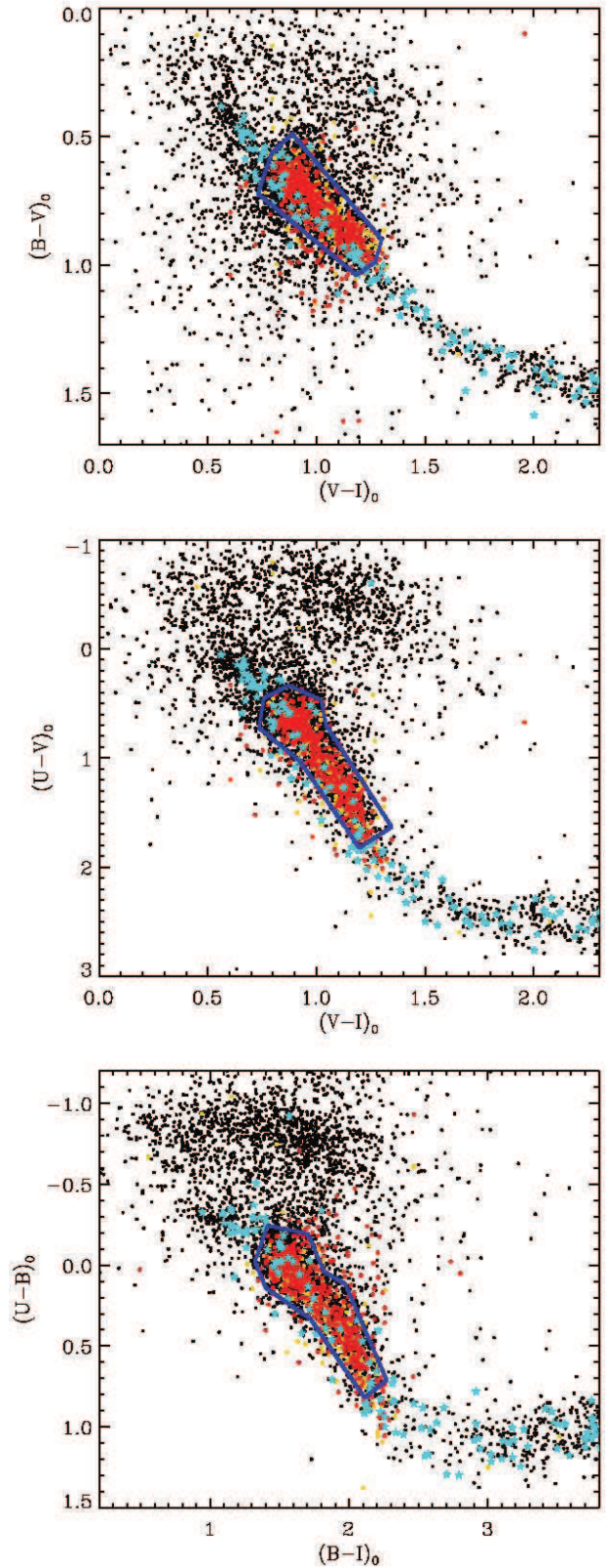}
\caption{Color selection of GC candidates in NGC 1399. 
The selection criteria demarcated by blue lines were determined 
based on confirmed GCs with radial velocity measurements (red dots;  \citealt{Schuberth10})   
and/or with size measurements (orange dots; derived from $HST$ ACS data). 
The cyan asterisks represent foreground stars confirmed with radial velocity measurements  \citealt{Schuberth10}).  
GC candidates are clearly separated from background sources in the $(U-B)_{0}$ color.  
\label{fig6_twocol}}
\end{figure}

We also investigated the radial variations of completeness around 
NGC 1399, NGC 1404, and NGC 1387. In the central region of a galaxy, 
the detection effectiveness at a certain magnitude decreases as the 
surface brightness of the galaxy increases, and it finally becomes zero 
at a saturated region. To quantify this effect in our observations, 
we performed separate artificial star tests near the galaxy centers.
We added 8000 artificial stars on the individual CCD chips 
that contain the galaxies (NGC 1399, NGC 1404, and NGC 1387), 
this time with a centrally concentrated radial distribution centered on each galaxy. 
This experiment was repeated five times with the positions of artificial stars altering each time.
The completeness was then calculated as a function of 
magnitude in concentric circles with radial intervals of 0\farcm2.
Figure~\ref{fig4_Com_radial} shows the $V$-band completeness functions at various radial   
bins from the center of each galaxy. The thick black line represents 
the completeness function for the overall region, as derived above. 
As expected from the strong background lights near the centers of our target galaxies,  
we find that the completeness is very low near the center. 
The completeness increases as one moves out from the center, 
and then flattens out beyond 1\farcm3 for NGC 1399, 
0\farcm7 for NGC 1404, and 0\farcm5 for NGC 1387.

\section{CONSTRUCTION OF GLOBULAR CLUSTER CATALOGS}

\subsection{Selection of Globular Cluster Candidates}

At the distance of the Fornax cluster \citep[$\sim$\,20.0 Mpc;][]{Blakeslee09}, 
1 pc is equal to 0\farcs01, so a typical GC appears as a point-source 
in our Mosaic images. We therefore can not select GCs based on their 
apparent sizes but must rely on photometric colors. Details on the 
selection process of GC candidates are given below.

We start by using the ALLFRAME parameters CHI and SHARP to select 
point-sources from our photometric catalog. Figure~\ref{fig5_chi} shows the 
SHARP, CHI, and the ALLFRAME photometric error as a function of 
$U_{0}$ magnitude with solid lines showing the selection criteria.    
While the figure only shows the $U$-band selection, we applied the 
same selection criteria to all bands simultaneously, and the resulting 
selection includes 5500 point-source objects. We then employ color 
cuts using our multi band photometry as demonstrated in Figure~\ref{fig6_twocol} to 
select GC candidates. The blue selection boxes shown in the three 
color--color diagrams were determined on the basis of the findings below. 
First, we matched our point-source objects with the spectroscopic catalog 
of \citet{Schuberth10} and found that the NGC 1399 GCs, as identified 
by their measured line-of-sight velocities, form a tight sequence (red 
dots in Figure~\ref{fig6_twocol}) in the color--color diagrams, whereas foreground stars 
are located along an extended sequence (cyan dots in Figure~\ref{fig6_twocol}). 
Second, we downloaded $Hubble~Space~Telescope~(HST)$/ACS 
images of NGC 1399 from the archive and identified genuine GCs based on 
their sizes. The sizes were measured using ISHAPE \citep{Larsen99} and 
following the same procedure outlined in \citet{Strader06}. We then 
cross-identified the GCs with our point-source objects, and found that 
most of them lie inside the tight sequence (orange dots in Figure~\ref{fig6_twocol}).   
To select GC candidates, we further applied a magnitude cut of 
$V_{0} \leq 20.6$ (corresponding to $M_{V} \lesssim -11$) to the   
color--color-selected sample. This magnitude limit was determined 
based on the brightest GC in the spectroscopic catalog of 
\citet{Schuberth10}, and we note that it is consistent with the 
limit used by \citet{Mieske06} to separate GCs from their ultra 
compact dwarf galaxies sample. The color--color and magnitude 
cuts result in a final sample of 2037 GC candidates.

\begin{figure*}
\epsscale{1.2}
\plotone{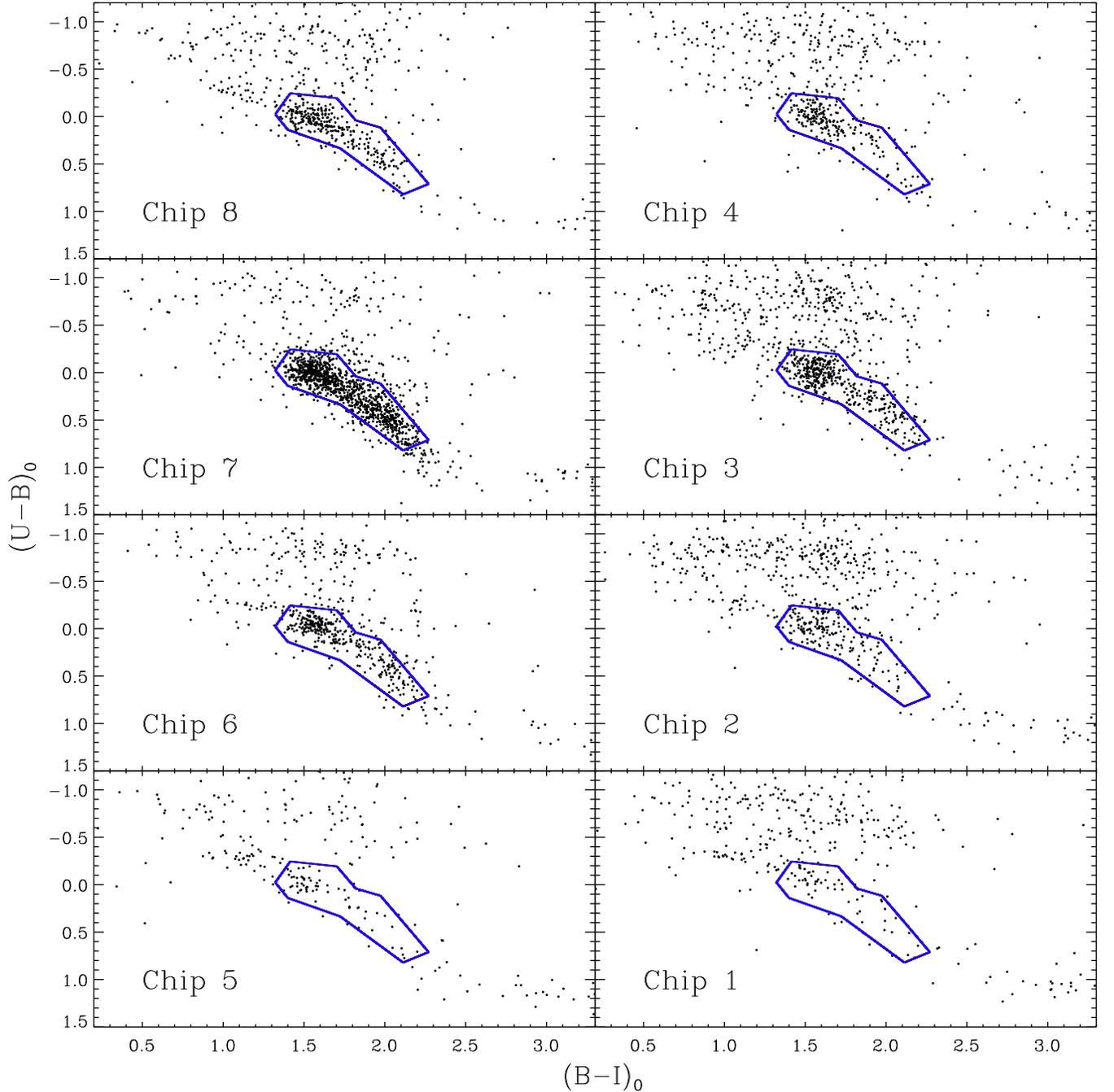}
\caption{Color--color diagrams of point-sources on each chip. 
Solid boxes indicate the selection criteria for the GC candidates and 
the chip numbers are indicated in the lower-left corner of each panel. 
Chips 7, 6, and 3 contain NGC 1399, NGC 1404, and NGC 1387, respectively.
Chip 1, which is located farthest away from the galaxies,
is used to estimate the upper limit of the contamination level of our data. \\
\label{fig7_ctoc}}
\end{figure*}

Figure~\ref{fig6_twocol} demonstrates the advantage of using $U$ band for separating 
star-forming background galaxies from the extragalactic GCs.
The background galaxies have $(B-V)_{0}$ colors similar to the bluer GCs,  
but due to their star-forming nature, galaxies exhibit significantly 
bluer colors in $(U-V)_{0}$ and $(U-B)_{0}$ than the GCs.   
This allows us to effectively reduce contamination by background galaxies.
The same effect was found by \citet{Dirsch03} using the $C-T1$ Washington color index.  
Nonetheless, our selection box still contains several foreground 
stars, and it is important to quantify the contamination level of 
our color--color-selected objects. We provide the estimation of the 
contamination level below.

\begin{figure*}
\epsscale{0.95}
\plotone{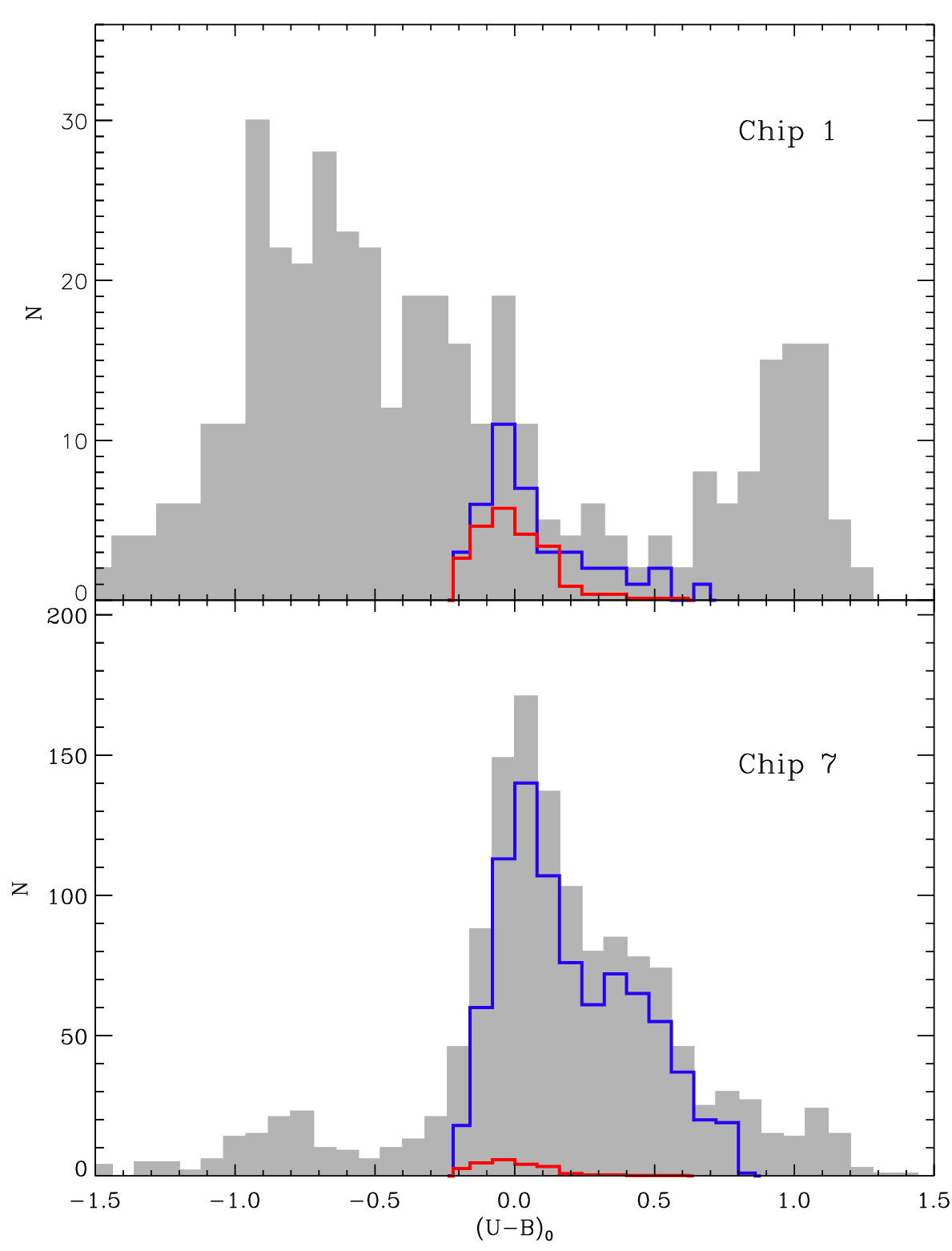}
\caption{$U-B$ color distributions for all point-sources (shaded histogram), 
GC candidates (blue), and synthetic stars from Besan{\c c}on model (red) in chip 1 and chip 7.
The similarities found in the red and blue color distributions in the upper panel
indicate that chip 1 is a good control field. \\
\label{fig8_Colbe}}
\end{figure*}

\subsection{Estimation of Contamination}

To estimate the number of contaminating sources (foreground stars and 
background galaxies) in our GC candidate list, it would be best to 
use a control field observed near the science field with the 
same observational conditions. However, this approach is impractical 
for us because the $U$-band observations require very long exposure 
times. As an alternative, we used two independent methods to estimate 
the contamination level of our GC candidate list.

\begin{figure*}
\epsscale{1.2}
\plotone{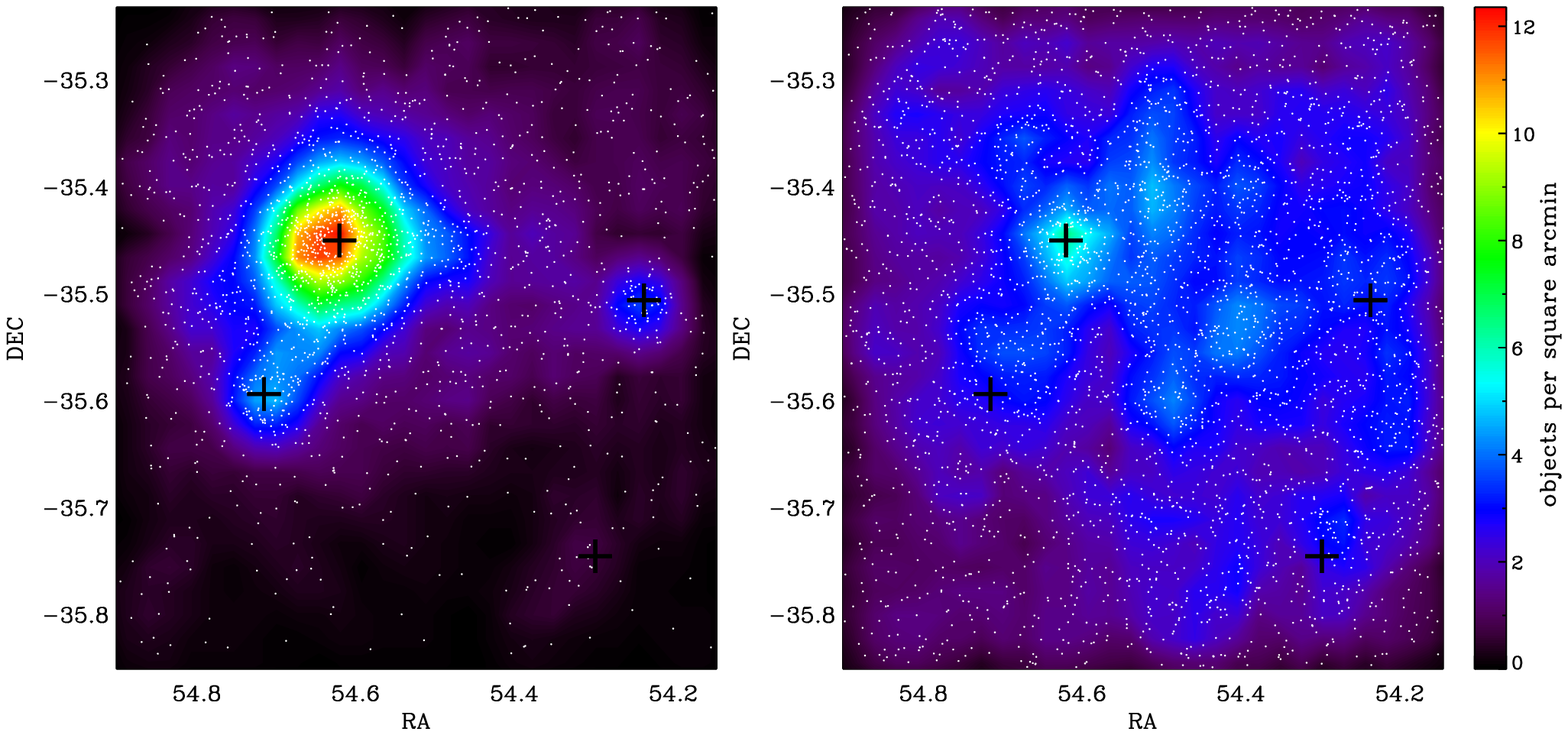}
\caption{Spatial distribution of GC candidates (left) and foreground/background objects (right) 
over color-filled density contour maps with the color scale from black to red for increasing surface number density. 
Black crosses represent the positions of the optical centers of galaxies (from top, NGC 1399, NGC 1387, NGC 1404, and NGC 1389).
\label{fig9_cont}}
\end{figure*}

\begin{deluxetable*}{ccccccccccc}
\tabletypesize{\scriptsize}
\tablecolumns{11}
\tablecaption{$UBVI$ Photometry of GC Candidates in the Fornax Galaxy Cluster}
\tablewidth{0pt}
\tablehead{\colhead{GC ID} & {R.A. (J2000)} & {Decl. (J2000)} & {$U_{0}$} & {$U$ Error} & {$B_{0}$} & {$B$ Error}  & {$V_{0}$} & {$V$ Error}  & {$I_{0}$} & {$I$ Error}}
\startdata
1 &  54.4670781 & --35.8412594 & 23.863 & 0.057 & 23.466 & 0.018 & 22.611 & 0.025 & 21.614 & 0.018 \\
2 &  54.4048663 & --35.8368808 & 23.099 & 0.028 & 22.941 & 0.011 & 22.069 & 0.015 & 21.119 & 0.010 \\
3 &  54.1532595 & --35.8313662 & 23.048 & 0.038 & 22.743 & 0.012 & 21.970 & 0.015 & 20.920 & 0.020 \\
4 &  54.3887923 & --35.8307234 & 24.085 & 0.070 & 23.983 & 0.029 & 23.384 & 0.049 & 22.462 & 0.030 \\
5 &  54.3871824 & --35.8300091 & 23.772 & 0.054 & 23.865 & 0.025 & 23.147 & 0.038 & 22.327 & 0.028 \\
6 &  54.3545131 & --35.8284302 & 23.147 & 0.030 & 22.633 & 0.008 & 21.661 & 0.011 & 20.490 & 0.006 \\
7 &  54.5147090 & --35.8246663 & 22.822 & 0.023 & 22.949 & 0.014 & 22.247 & 0.015 & 21.430 & 0.013 \\
8 &  54.3057038 & --35.8223111 & 24.007 & 0.061 & 24.027 & 0.028 & 23.363 & 0.045 & 22.429 & 0.028 \\
9 &  54.2008513 & --35.8137756 & 24.064 & 0.070 & 24.111 & 0.030 & 23.383 & 0.045 & 22.486 & 0.030 \\
10 &  54.4062928 & --35.8090401 & 21.291 & 0.007 & 21.338 & 0.004 & 20.674 & 0.005 & 19.854 & 0.003 \\
\nodata & \nodata & \nodata & \nodata & \nodata & \nodata & \nodata & \nodata & \nodata & \nodata & \nodata \\
\nodata & \nodata & \nodata & \nodata & \nodata & \nodata & \nodata & \nodata & \nodata & \nodata & \nodata \\
\nodata & \nodata & \nodata & \nodata & \nodata & \nodata & \nodata & \nodata & \nodata & \nodata & \nodata
\enddata
\tablenotetext{\,}{$\bf{Note.}$ The first 10 rows of the table are reproduced here; the full version of this table containing the 2,037 GC candidates is available online at the CDS.\\}
\label{tbl-cat}
\end{deluxetable*}

On average, chip 1 is located farthest away from the target galaxies 
NGC 1399, NGC 1387, and NGC 1404, so it has the least number of GCs 
expected among the eight CCD chips.\footnote{NGC 1389 is known to have 
very few GCs (see Section~4.1), so we only considered the other 
three Fornax galaxies here.} In Figure~\ref{fig7_ctoc}, we compare the $(U-B)_{0}$ versus 
$(B-I)_{0}$ color--color diagrams of the 5500 point-source objects for each  
chip. In chip 1, there are 41 objects that are selected as GC 
candidates using the color--color selection box as in Figure~\ref{fig6_twocol}. If we 
assume that all 41 objects are contaminants, this implies that our 
catalog of GC candidates has a contamination level of 16\,\%. 
We note that 16\,\% is in fact an upper limit because chip 1, which is 
at an average distance of 24\arcmin\,($\sim$\,140 kpc) from the center 
of NGC 1399, is likely to contain several GCs given that the NGC 1399 
GCs are known to extend out to a projected distance of at least 45\arcmin\,
($\sim$\,260 kpc; \citealt{Bassino06}).

We further estimated the number of foreground stars using the 
Besan{\c c}on model of the Milky Way population \citep{Robin03}.
We generated a synthetic catalog of stars in the direction of the sky 
consistent with our target field and then selected stars using the 
same color--color cut as used in Figure~\ref{fig6_twocol}. The estimated number of 
foreground stars in our GC candidate list turned out to be 180   
and the corresponding contamination level is 9\,\%. As expected, this  
is lower than the upper limit set by using chip 1 as the control 
field above. Based on the two estimates, the contamination level 
of our GC candidate list is likely to be in the 9\,\%\,--16\,\% range.

Although the contamination level itself may not be that high, it is 
important to test whether the contaminating sources will affect our 
GC color distribution analyses. Figure~\ref{fig8_Colbe} shows the comparison of  
$(U-B)_{0}$ color distributions for all point-sources (shaded histogram),  
our GC candidate list (blue), and synthetic stars from the 
Besan{\c c}on model (red) between chips 7 and 1.
Based on the similarities found in the red and blue color distributions
in the upper panel of Figure~\ref{fig8_Colbe}, chip 1 seems to be a good control field.
From the same figure, we find that the foreground MW stars are expected 
to have a $(U-B)_{0}$ color distributions that is peaked toward the bluer   
side similar to the $(U-B)_{0}$ color distribution of GC candidates.   
However, as demonstrated in the lower panel of Figure~\ref{fig8_Colbe}, the number of 
these foreground stars, when compared to that of the GC candidates, 
is too small to have any impact on our color distribution analyses 
presented in Sections~4.3 and 4.4.

\begin{figure*}
\epsscale{0.8}
\plotone{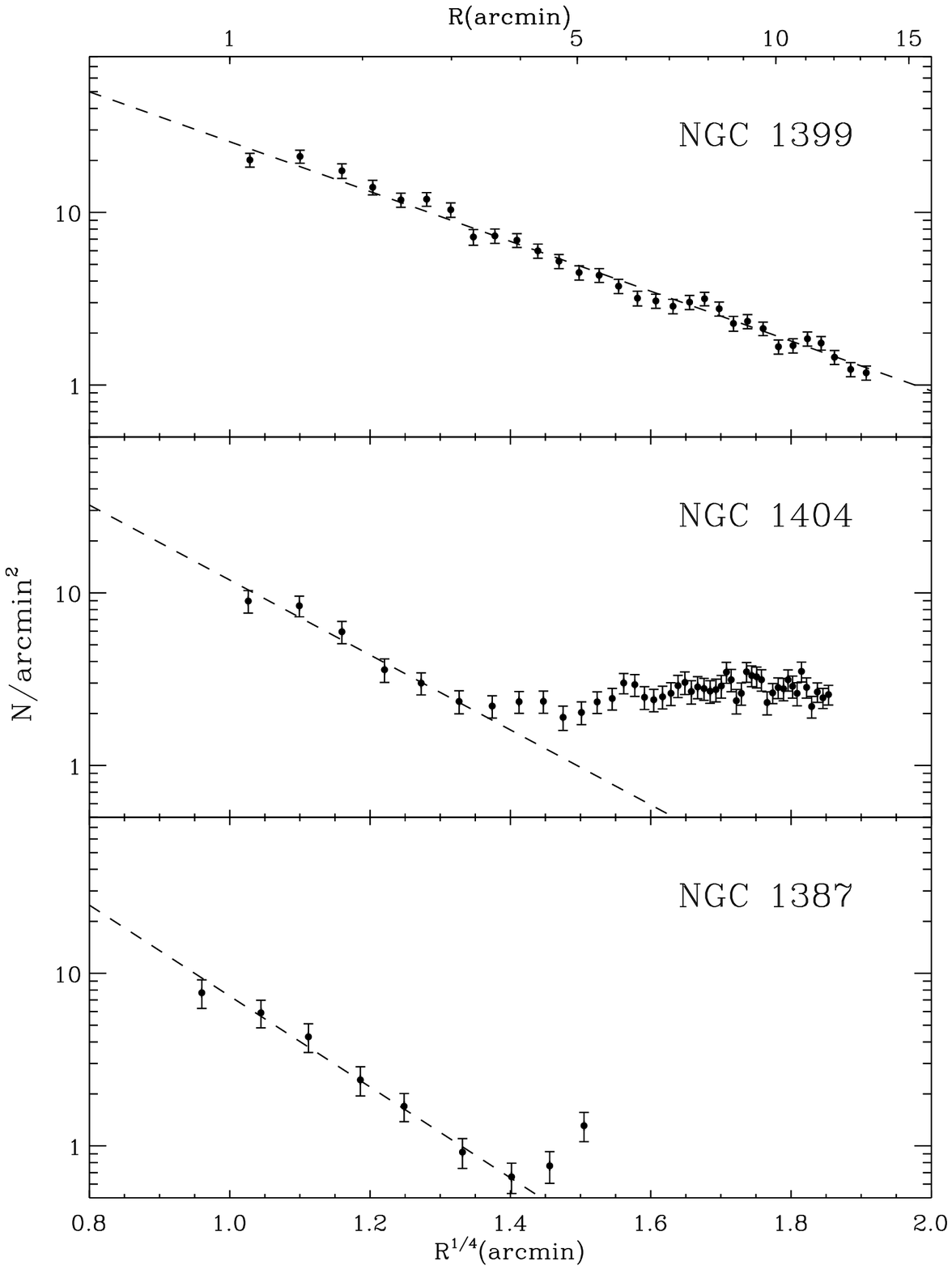}
\caption{Number of GC candidates per unit area projected on the sky as a function of radial distance from the center of the galaxy. 
Radial bins are defined to contain approximately equal numbers of GCs 
($\sim$\,100 GCs for NGC 1399; $\sim$\,40 GCs for NGC 1404; $\sim$\,25 GCs for NGC 1387). 
The numbers are corrected for the spatial completeness, but not for the contamination by foreground/background objects (see the text).   
Dashed line shows de Vaucouleurs' $R^{\rm \,1/4}$ fit of each profile.\\
\label{fig10_RDen}}
\end{figure*}

\subsection{Photometric Catalog}

Table~\ref{tbl-cat} presents the photometric catalog of total 2,037 GC candidates. 
The columns give the following information.\\
\noindent $Column~\textit{1}$.~Identification number \\
\noindent $Columns~\textit{2~and~3}$.~Right ascension and declination (J2000) \\
\noindent $Columns~\textit{4~and~5}$.~$U_{0}$ magnitude and its rms uncertainty\\   
\noindent $Columns~\textit{6~and~7}$.~$B_{0}$ magnitude and its rms uncertainty\\   
\noindent $Columns~\textit{8~and~9}$.~$V_{0}$ magnitude and its rms uncertainty\\   
\noindent $Columns~\textit{10~and~11}$.~$I_{0}$ magnitude and its rms uncertainty.\\  
\indent We note that all magnitudes in the catalog are corrected for Galactic extinction.

\begin{figure*}
\epsscale{1.1}
\plotone{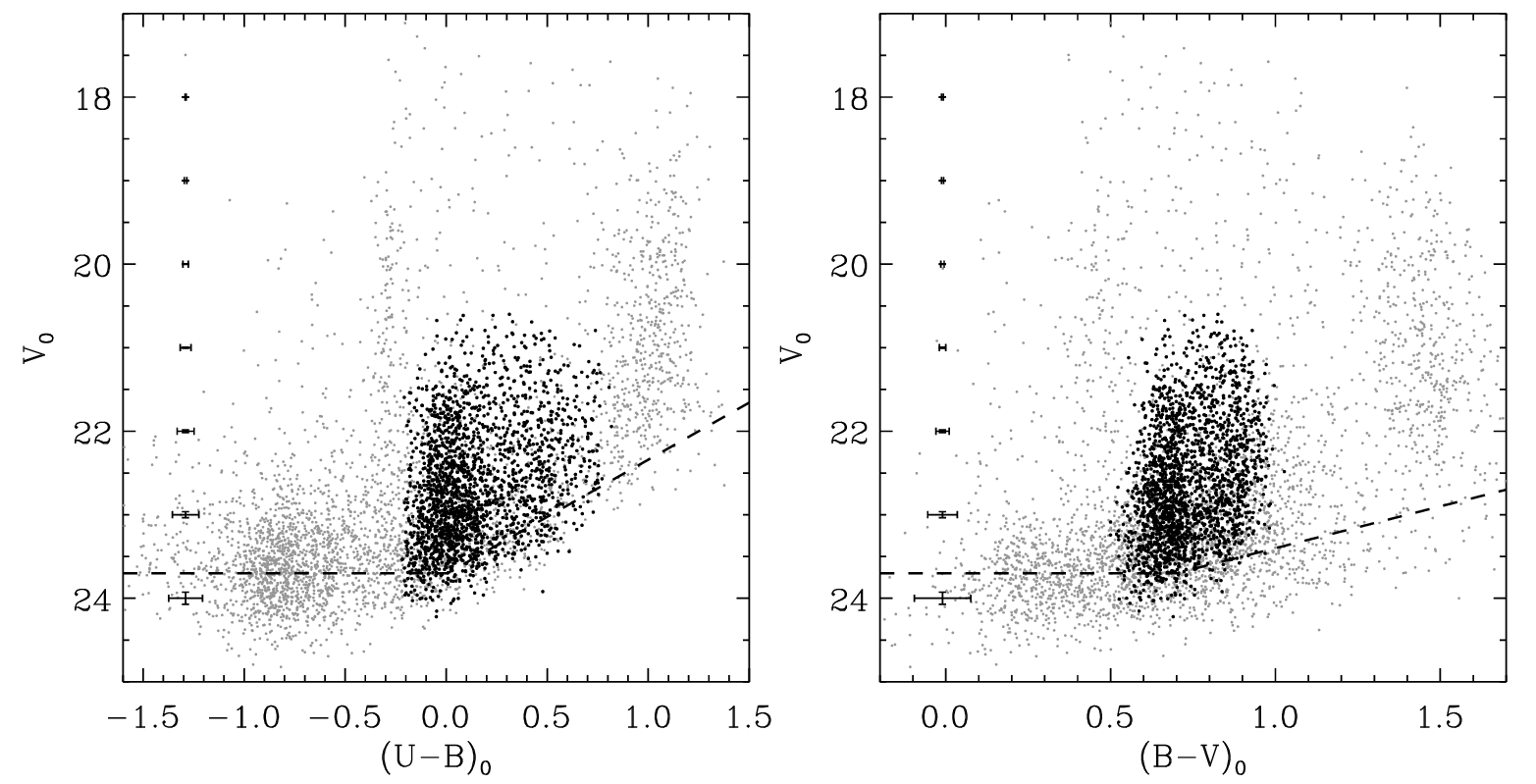}
\caption{Color--magnitude diagrams of point-sources. 
Black dots represent GC candidates, and gray dots represent foreground/background objects. 
Observational uncertainties as a function of $V_{0}$ are shown by error bars in the left corner of each diagram.
Dashed lines indicate the limiting magnitude. The tilt of the limiting magnitude 
results from lower detection efficiency of red GCs than blue GCs in the $U$ band.
The GC candidates fainter than the limiting magnitude are taken out from the subsequent analysis. \\
\label{fig11_CMD}}
\end{figure*}

\begin{deluxetable*}{l c c c c c c c c c}
\tabletypesize{\scriptsize}
\tablecolumns{10}
\tablecaption{Results from GMM Analysis for Color Distributions of GCs}
\tablewidth{0pt}
\tablehead{\colhead{Color} & {$\mu_{b}$} & {$\sigma_{b}$} & {$\mu_{r}$} & {$\sigma_{r}$} & {$f_{r}$} & {$\sigma_{r}$}/{$\sigma_{b}$} & {$p(\chi^{2})$} & {$p(DD)$} & {$p(kurt)$}}
\startdata
\multicolumn{10}{c}{NGC 1399}\\
\hline \\
 $U-B$ &  0.016 $\pm$ 0.006 & 0.102 $\pm$ 0.003 & 0.372 $\pm$ 0.021 & 0.188 $\pm$ 0.009 & 0.34 & 1.84 & 0.010 & 0.090 & 0.860 \\
 $U-V$ &  0.670 $\pm$ 0.007 & 0.113 $\pm$ 0.005 & 1.147 $\pm$ 0.022 & 0.269 $\pm$ 0.008 & 0.42 & 2.38 & 0.010 & 0.100 & 0.140 \\
 $U-I$ &  1.557 $\pm$ 0.008 & 0.134 $\pm$ 0.006 & 2.188 $\pm$ 0.025 & 0.357 $\pm$ 0.008 & 0.45 & 2.66 & 0.010 & 0.080 & 0.010 \\
 $B-V$ &  0.673 $\pm$ 0.004 & 0.058 $\pm$ 0.002 & 0.849 $\pm$ 0.007 & 0.060 $\pm$ 0.003 & 0.34 & 1.03 & 0.010 & 0.070 & 0.010 \\
 $B-I$ &  1.571 $\pm$ 0.005 & 0.087 $\pm$ 0.003 & 1.925 $\pm$ 0.014 & 0.135 $\pm$ 0.007 & 0.37 & 1.55 & 0.010 & 0.050 & 0.010 \\
 $V-I$ &  0.910 $\pm$ 0.005 & 0.063 $\pm$ 0.003 & 1.108 $\pm$ 0.013 & 0.068 $\pm$ 0.007 & 0.30 & 1.08 & 0.010 & 0.060 & 0.010 \\
\cutinhead{NGC1404}\\
 $U-B$ & --0.004 $\pm$ 0.012 & 0.079 $\pm$ 0.008 & 0.490 $\pm$ 0.024 & 0.120 $\pm$ 0.018 & 0.39 & 1.52 & 0.001 & 0.017 & 0.001 \\
 $U-V$ &  0.682 $\pm$ 0.015 & 0.101 $\pm$ 0.011 & 1.393 $\pm$ 0.026 & 0.136 $\pm$ 0.017 & 0.39 & 1.35 & 0.001 & 0.001 & 0.001 \\
 $U-I$ &  1.631 $\pm$ 0.018 & 0.124 $\pm$ 0.015 & 2.583 $\pm$ 0.029 & 0.161 $\pm$ 0.019 & 0.39 & 1.30 & 0.001 & 0.001 & 0.001 \\
 $B-V$ &  0.686 $\pm$ 0.007 & 0.052 $\pm$ 0.007 & 0.901 $\pm$ 0.005 & 0.029 $\pm$ 0.004 & 0.38 & 0.56 & 0.001 & 0.014 & 0.001 \\
 $B-I$ &  1.634 $\pm$ 0.011 & 0.075 $\pm$ 0.011 & 2.091 $\pm$ 0.009 & 0.051 $\pm$ 0.006 & 0.39 & 0.68 & 0.001 & 0.001 & 0.001 \\
 $V-I$ &  0.949 $\pm$ 0.006 & 0.041 $\pm$ 0.006 & 1.190 $\pm$ 0.006 & 0.030 $\pm$ 0.005 & 0.39 & 0.73 & 0.001 & 0.001 & 0.001 \\
\cutinhead{NGC1387}\\
 $U-B$ & --0.019 $\pm$ 0.017 & 0.084 $\pm$ 0.011 & 0.446 $\pm$ 0.030 & 0.128 $\pm$ 0.017 & 0.49 & 1.52 & 0.001 & 0.041 & 0.001 \\
 $U-V$ &  0.693 $\pm$ 0.020 & 0.095 $\pm$ 0.013 & 1.355 $\pm$ 0.035 & 0.160 $\pm$ 0.021 & 0.49 & 1.68 & 0.001 & 0.011 & 0.001 \\
 $U-I$ &  1.624 $\pm$ 0.023 & 0.114 $\pm$ 0.014 & 2.505 $\pm$ 0.044 & 0.196 $\pm$ 0.022 & 0.49 & 1.72 & 0.001 & 0.005 & 0.001 \\
 $B-V$ &  0.707 $\pm$ 0.010 & 0.046 $\pm$ 0.008 & 0.904 $\pm$ 0.010 & 0.045 $\pm$ 0.008 & 0.51 & 0.98 & 0.001 & 0.039 & 0.001 \\
 $B-I$ &  1.642 $\pm$ 0.012 & 0.058 $\pm$ 0.006 & 2.056 $\pm$ 0.017 & 0.079 $\pm$ 0.009 & 0.49 & 1.36 & 0.001 & 0.002 & 0.001 \\
 $V-I$ &  0.931 $\pm$ 0.007 & 0.038 $\pm$ 0.009 & 1.148 $\pm$ 0.013 & 0.052 $\pm$ 0.009 & 0.49 & 1.37 & 0.001 & 0.017 & 0.001 \\
\enddata
\label{tbl-GMM}
\end{deluxetable*}

\begin{figure*}
\epsscale{1.0}
\plotone{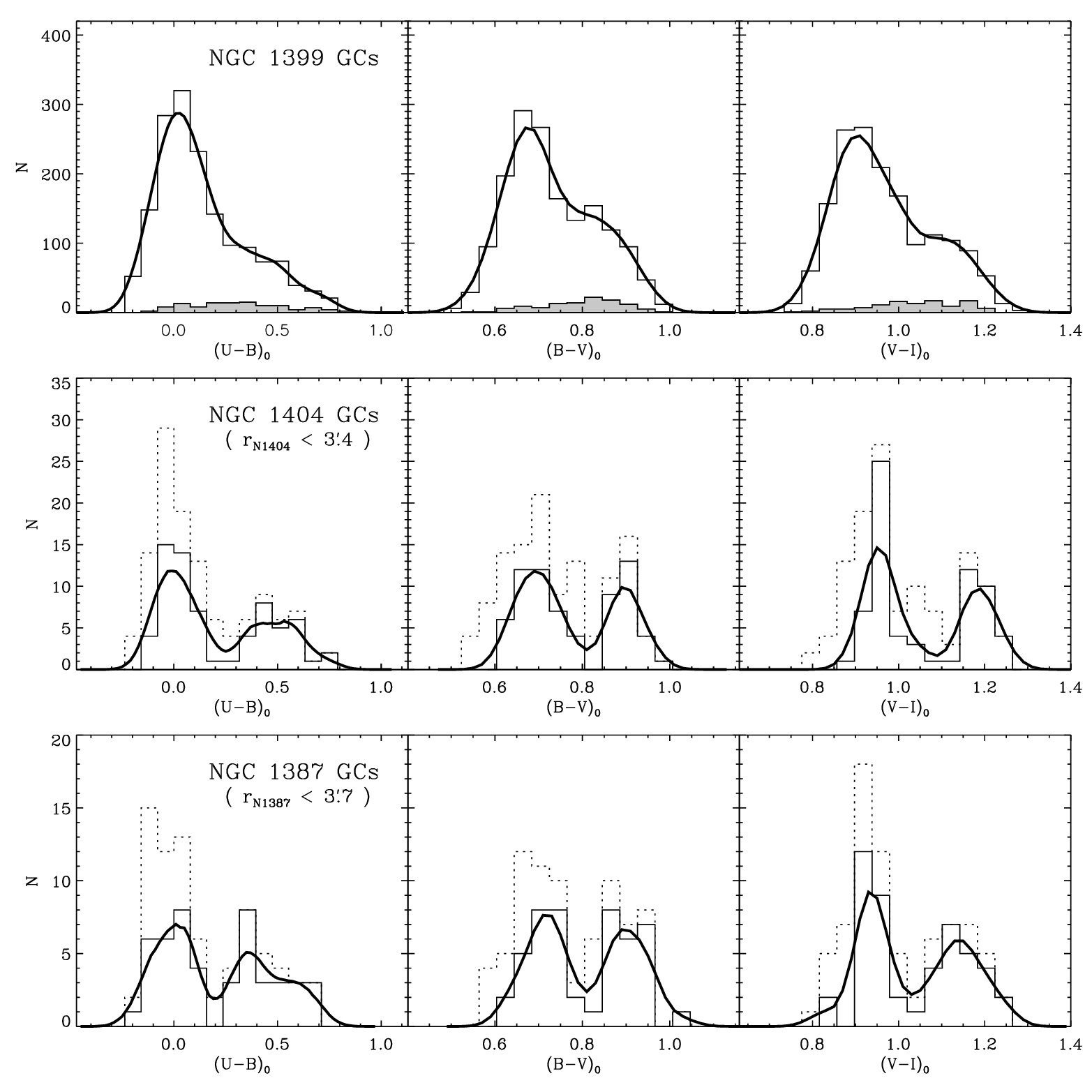}
\caption{Color distributions of GCs in NGC 1399, NGC 1404, and NGC 1387. 
Thick solid lines are smoothed histograms with Gaussian kernels 
of $\sigma$($U-B$) = 0.08, $\sigma$($B-V$) = 0.04, and $\sigma$($V-I$) = 0.04, respectively. 
The $\sigma$ values are set to be comparable to the mean errors in each color. 
Filled gray histograms in the top panels show the color distributions of the brightest GCs ($V_{0}<21.5$). 
Dotted histograms in the middle and bottom panels represent the color histograms 
of entire set of GCs within the extent of each galaxy, 
and solid histograms show the ``corrected'' distributions for the contamination by NGC 1399 GCs (see the text). \\
\label{fig12_Colhist}} 
\end{figure*}

\section{RESULTS AND DISCUSSION}

\subsection{Spatial Distribution}

\begin{figure*}
\epsscale{1.0}
\plotone{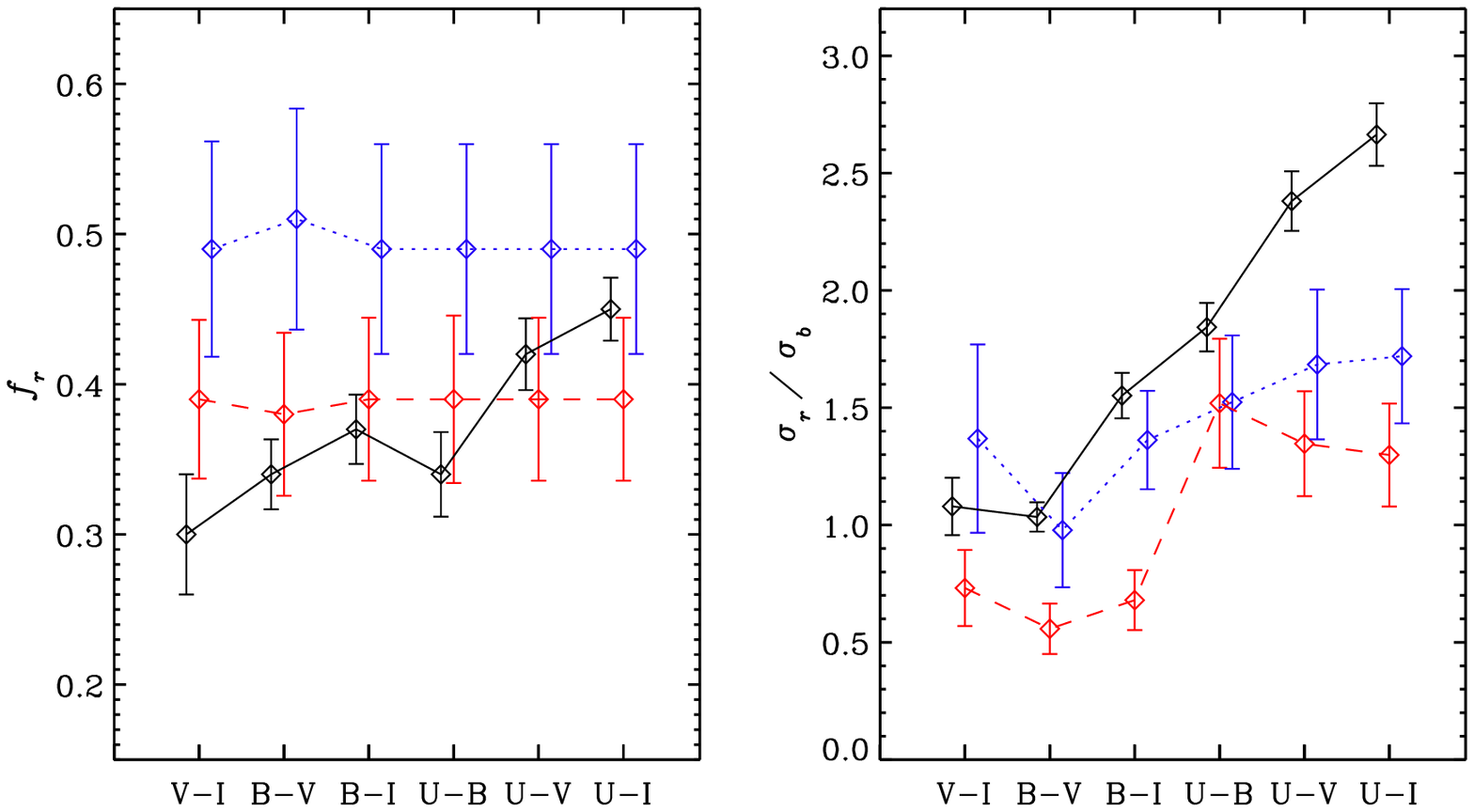}
\caption{Variation in the red GC fractions (left) and 
the ratios of standard deviations between blue and red GC colors (right) for various colors. 
Black solid lines, red dashed lines, and blue dotted lines are for NGC 1399, NGC 1404, and NGC 1387, respectively. 
For NGC 1404 and NGC 1387, the values 
were derived from the corrected color distributions for the contribution of NGC 1399 GCs in Figure~\ref{fig12_Colhist}.\\
\label{fig13_Tab4}} 
\end{figure*}

Figure~\ref{fig9_cont} maps the spatial distributions of GC candidates (left panel) 
and foreground/background objects (right). 
The color-filled contours show the distribution of surface number density. 
The black crosses mark the positions of the optical centers of galaxies
(from top, NGC 1399, NGC 1387, NGC 1404, and NGC 1389). 
In the left panel, three GC systems  are clearly visible 
around NGC 1399, NGC 1404, and NGC 1387.
The number of GCs around NGC 1389 is too small, 
so we omit the galaxy from the following analysis. 
In the right panel, the distribution of foreground/background objects appears fairly uniform. 
The slight overdensity in the region between NGC 1399 and NGC 1387
is likely due to the presence of a distant galaxy cluster,
given that many objects in this region are bluer than --\,0.3 in $(U-B)_{0}$. 
Another overdense region is right on the center of NGC 1399, 
which indicates that GCs in highly dense regions 
can be misclassified as foreground/background objects 
in our GC selection procedure, as mentioned in Section 3.
Note that the residuals of the mosaic CCD gaps are discernible 
as a vertical feature in the middle of the panel.

It is noteworthy that the GC distribution around NGC 1399 shows interesting features.
On the one hand, there is an overabundance of GCs in the outer region of NGC 1399 
toward NGC 1404 and NGC 1387. 
This was also pointed out by \citet{Bassino06},
and it may indicate interactions of NGC 1399 with NGC 1404 and NGC 1387 
in the recent past (\citealt{Forbes97, Bekki03}). 
On the other hand, 
a  $\sim$\,0\farcm5 displacement is found
between NGC 1399's optical center 
and the center of the GC distribution in the inner region of the galaxy
as denoted by a red contour in the density map. 
\citet{Paolillo02} reported an asymmetric X-ray halo for NGC 1399,
consisting of three components with different centers; a central component, 
a galactic component centered 1$\arcmin$ southwest of NGC 1399,
and a cluster component centered $\sim$\,5\farcm6 northeast of the galaxy.
While the center of the GC distribution does not coincide with the X-ray centers, 
the asymmetric distribution of the GCs may suggest 
that the NGC 1399 is not yet dynamically relaxed
and may be undergoing merger events.

Figure~\ref{fig10_RDen} presents the surface number densities 
against the galactocentric radius for the three galaxies.
In order to examine the radial extent of each GC system, 
we calculated the surface GC number density as a function of the distance from the galaxy center. 
We first set up a series of radial bins (annulus), each containing approximately equal numbers of GCs 
($\sim$\,100 GCs for NGC 1399; $\sim$\,40 GCs for NGC 1404; $\sim$\,25 GCs for NGC 1387). 
The number of GCs in each bin was then corrected using the corresponding completeness function of the bin.
We finally divided the corrected numbers of GCs by the area of annuli 
to get surface densities. 
The surface density profile of each galaxy 
is well described by de Vaucouleurs' $R^{\rm \,1/4}$ fits (dashed lines). 
We determine the limiting radii of GC systems of NGC 1404 (3\farcm4) and NGC 1387 (3\farcm7) 
as the radial points where the density profiles begin to depart from de Vaucouleurs' law. 
Since the GC system of NGC 1399 is extended ($\sim$\,45\arcmin; \citealt{Bassino06}) 
well beyond the field of view ($36\arcmin\times36\arcmin$) of our Mosaic observations, 
we regard the entire GCs in the images as NGC 1399's GC system,
except for the regions inside the limiting radii of NGC 1404 and NGC 1387.

Figure~\ref{fig11_CMD} presents the CMDs for all point-sources. 
Magnitudes and colors were corrected for the Galactic extinction, as mentioned in Section 2.3. 
The black and gray dots represent the GC candidates and foreground/background objects, respectively. 
The dashed lines indicate the limiting magnitude of 50\,\% completeness. 
The tilt of the limiting magnitude lines results from 
lower detection efficiency of red GCs than blue GCs in the $U$ band. 
The $(U-B)_{0}$ colors corresponding to GCs, background galaxies, and
foreground stars can be determined both by the $(U-B)_{0}$ versus $(B-I)_{0}$ diagram (the bottom panel of Figure~\ref{fig6_twocol}) 
and by the $V_{0}$--$(U-B)_{0}$ CMD (the left panel of Figure~\ref{fig11_CMD}).  
GCs are placed at $-0.2<(U-B)_{0}<0.8$,    
while most background galaxies are bluer at $(U-B)_{0}$ $\simeq$ $-0.8$
and foreground stars are at $(U-B)_{0}=-0.3$ and 1.1 close to both ends of the GC range. 
The $V_{0}$--$(B-V)_{0}$ CMD (right panel) shows
that many of the background galaxies have similar colors to the GCs, 
making it difficult to discriminate GCs from galaxies in this color. 
The foreground star sequences, particularly on the red side, 
are more clearly separated from GCs than in $(U-B)_{0}$.
This comparison corroborates that 
the use of various color combinations including $U$-band colors 
provides a powerful tool for discriminating GCs from other objects.

\subsection{Color--Magnitude Diagrams}

It is interesting to note that the CMDs do not show bimodality                   
for GCs brighter than $V_{0} \sim 21.5$.
The disappearance was previously reported for NGC 1399 
(\citealt{Ostrov98, Dirsch03, Blakeslee12}), 
a few galaxies in the Virgo Cluster (\citealt{Mieske06, Strader06}), 
and other giant galaxies (\citealt{Harris06, Faifer11}). 
The unimodal distribution is often explained 
by merging of blue and red sequences at the highest luminosities 
as a result of the blue tilt phenomenon \citep{Harris06},
which is a trend for the blue GCs to have redder colors at higher luminosities.
However, \citet{Ostrov98} and \citet{Forte07} have reported that 
there is no blue tilt in the GC system of NGC 1399.
Interestingly, our data reveal that 
the number of blue GCs rapidly drops at the brightest magnitudes ($V_{0}$ $<$ 21.5).
This suggests that the unimodal distribution of brightest GCs 
is not the consequence of the blue tilt, 
but instead owing to the relative scarcity of blue, brightest GCs.

\subsection{Color Distributions}
Figure~\ref{fig12_Colhist} shows three representative color histograms of 
the entire GC system within the limiting radius   				
in NGC 1399 (top), NGC 1404 (middle), and NGC 1387 (bottom). 
The GCs fainter than the limiting magnitudes and/or   
the ones located in the innermost region of each galaxy,   
where the completeness test is unreliable (see Section 2.3),  
are taken out from the subsequent analysis.  
The thick solid lines are smoothed histograms with Gaussian kernels. 
The filled histograms in the top row are for the brightest GCs, 
as mentioned in the previous section.
For NGC 1399 (top row), GCs show bimodal distributions for all colors considered 
in agreement with previous findings in $C-T1$ (\citealt{Dirsch03, Bassino06, Forte07}).
For NGC 1404 (middle) and NGC 1387 (bottom), 
the dotted histograms are for all GCs within the limiting radii of the galaxies.
The solid histograms represent ``corrected'' distributions,
for which the contribution of NGC 1399 GCs is statistically subtracted
from the original distributions and will be used in the further analysis. 
NGC 1404 GCs exhibit pronounced bimodality 
with an evident dip between two groups in all colors, 
consistent with the previous studies for NGC 1404's inner GCs 
by \citet{Grillmair99} and \citet{Larsen01}.
The clear separation between blue and red GCs for NGC 1387 GCs 
is also consistent with what was found in $C-T1$ by \citet{Bassino06b}. 
Blue GCs are much less abundant in NGC 1404 and NGC 1387 than in NGC 1399. 

A closer scrutiny of Figure~\ref{fig12_Colhist} reveals that 
the exact morphologies of the GC color distributions 
change depending on colors considered. 
The most distinctive case is given by the $(U-B)_{0}$ distribution of NGC 1399.  
Compared to the $(B-V)_{0}$ and $(V-I)_{0}$ histograms,   
the $(U-B)_{0}$ distribution exhibits more prominent blue GC peak  
and have red GCs with weaker peak and larger dispersion in color.
The $(U-B)_{0}$ distribution shows spread in colors of red GCs that is nearly a factor of two larger than that of blue GCs,  
while the $(B-V)_{0}$ and $(V-I)_{0}$ histograms have similar variances for the red and blue GCs.   
The histograms for NGC 1404 and NGC 1387
also show changes in their shape depending on colors.

To be more quantitative in the bimodality analysis, 
we used the Gaussian Mixture Modeling (GMM) code by \citet{Muratov10}.    
Table~\ref{tbl-GMM} presents the results of the GMM analysis for six color combinations 
($(U-B)_{0}$, $(U-V)_{0}$, $(U-I)_{0}$, $(B-V)_{0}$, $(B-I)_{0}$, and $(V-I)_{0}$) for the three galaxies.  
It gives the color index, the mean ($\mu)$, and the standard deviation ($\sigma$) of blue and red GC colors, 
the red GC fraction ($f_{r}=N_{\textnormal{\tiny{red}}}/N_{\textnormal{\tiny{total}}}$), 
and the ratio of the standard deviations between blue and red GCs ($\sigma_r$/$\sigma_b$).
The last three columns summarize the probabilities 
of preferring a unimodal distribution over a bimodal distribution ($p$-values)
derived based on the likelihood ratio test (LRT; $\chi^{2}$), 
on the separation of the means relative to their variances ($DD$) 
and on the kurtosis of a distribution ($kurt$).\footnote{\footnotesize 
Note that the kurtosis gives a necessary (but not sufficient) condition of bimodality, 
and it can be used only as an additional check of other statistics \citep{Muratov10}.
For all galaxies, $p(\chi^{2})$ from the LRT is $\le0.01$ in all colors, 
indicating the bimodal distributions are preferred over the unimodal ones. 
The $p(DD)$ values lead to the same conclusion in all cases. 
The $p(kurt)$ values support the bimodal distributions in most cases, 
but favor a unimodal distribution for $(U-B)_{0}$ of NGC 1399 with $p(kurt)=0.860$.}  

Figure~\ref{fig13_Tab4} presents the results from the GMM analysis for all six color combinations. 
To quantify the shape of histograms, we use the red GC fractions (left) 
and the ratios of standard deviations between blue and red GC colors (right).
NGC 1399 GC sample (black solid lines) does not include GCs around NGC 1404 and NGC 1387,
and the GC sample of NGC 1404 (red dashed lines) and NGC 1387 (blue dotted) is 
corrected for NGC 1399 GC contribution.
For NGC 1399, we find that both the numbers and dispersions of blue and red GCs 
change significantly depending on colors that are used. 
The trend is more evident when comparing the $U$-band colors with the rest.  
For NGC 1404 and NGC 1387,
the interesting features emerge after the correction for the contribution of NGC 1399 GCs. 
First, the red GC fractions in all colors remain constant regardless of the colors,
as expected in Figure~\ref{fig12_Colhist} showing distinct separation between blue and red GCs.  
It is plausible that the early interactions of the galaxies with NGC 1399 have preferentially left the central, red GCs
and later blue GCs have been accreted from outer region of NGC 1399. 
Second, the color dispersions vary significantly depending on colors, 
following a similar pattern to the case of NGC 1399 for $V-I$, $B-V$, and $B-I$.
For $U$-band colors, however, the patterns are different        
in the sense that they tend to be constant within the errors.   
The variation of dispersion as a function of colors 
is not expected if the metallicity-to-color converting relations are straight.
Instead, the slopes seem different between the blue and red parts of the metallicity--color relations
and such an effect varies systematically from color to color 
(\citealt{Yoon06, Yoon11a, Yoon11b, Yoon12, Cantiello07, Santos12, Blakeslee12}).

\begin{figure*}
\epsscale{1.00}
\plotone{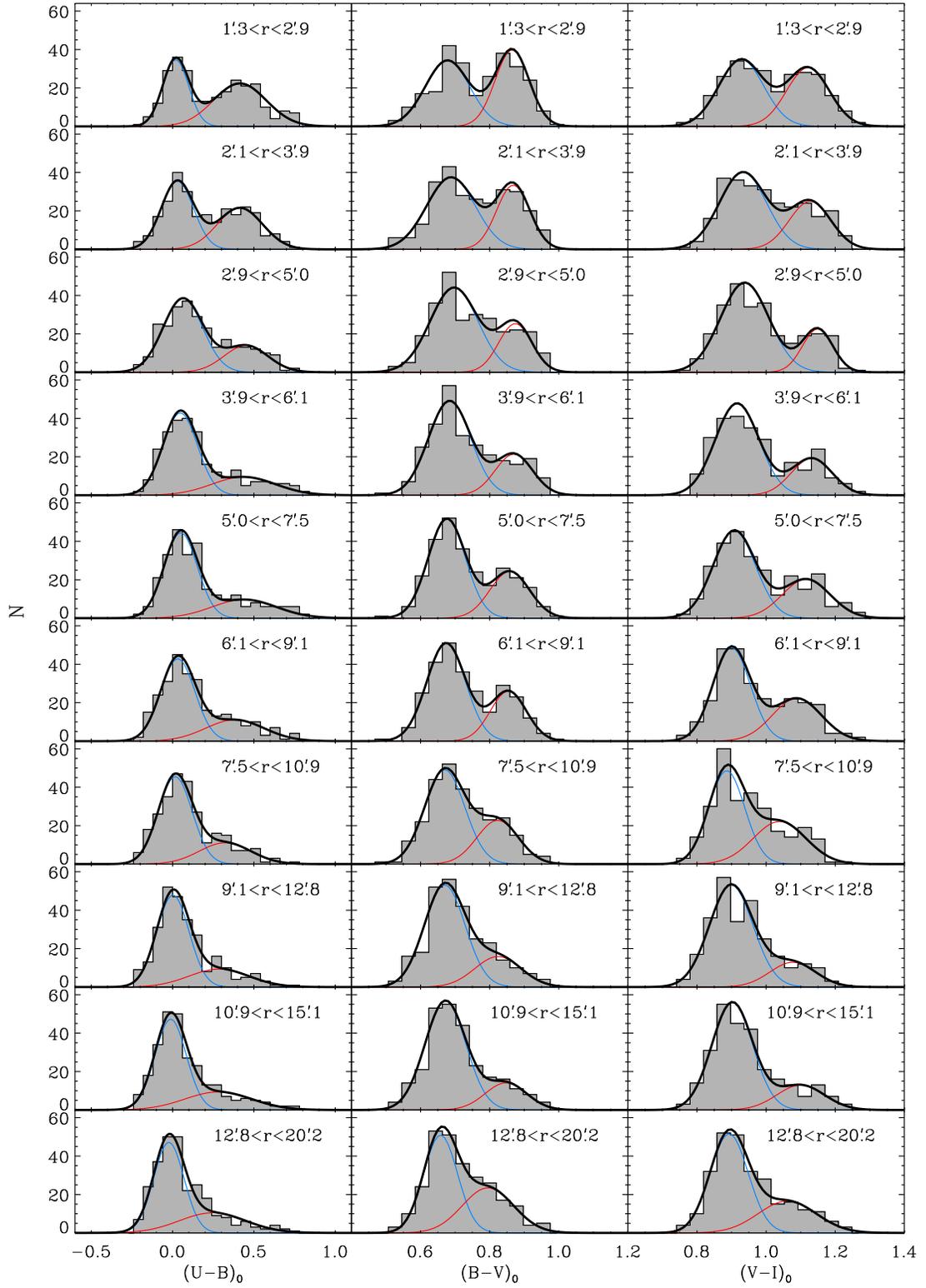}
\caption{
Histograms of $(U-B)_{0}$, $(B-V)_{0}$, and $(V-I)_{0}$ colors (left, middle, and right columns) for NGC 1399 GCs 
in various radial bins indicated in the upper-right corner within each panel.
The radial bins are defined to contain approximately equal numbers ($\sim$\,300) of GCs. 
The histograms are expressed by two (i.e., blue and red) Gaussian normal distributions based on GMM analysis. 
The blue, red, and black lines represent blue GCs, red GCs, and their sum, respectively.
\label{fig14_N1399h}} 
\end{figure*}

\begin{figure*}
\epsscale{1.0}
\plotone{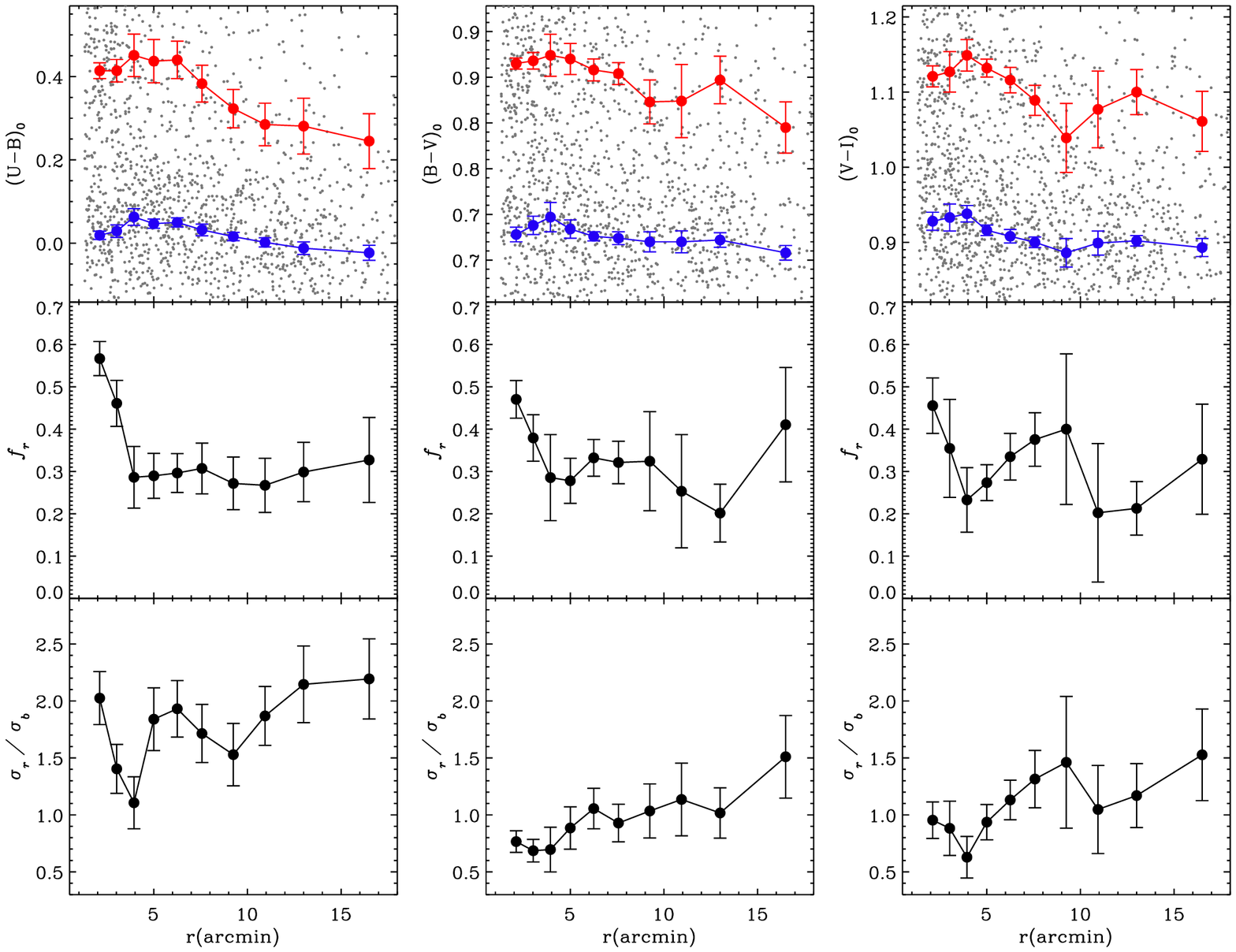}
\caption{Parameters of bimodality in $(U-B)_{0}$, $(B-V)_{0}$, and $(V-I)_{0}$ colors (left, middle, and right columns) 
as a function of galactocentric distance for NGC 1399 GCs. 
The top, middle, and bottom panels show the mean colors ($\mu_b$ and $\mu_r$ for blue and red GCs), 
the red GC fraction ($f_r$), and the ratios of color dispersions between blue and red GCs ($\sigma_r$/$\sigma_b$), 
respectively. \\
\label{fig15_N1399r}} 
\end{figure*}

\subsection{Radial Variations of Color Bimodality}

Figure~\ref{fig14_N1399h} displays the color distributions
as a function of galactocentric radius for GCs in NGC 1399.  
The subsequent analysis is only valid for NGC 1399 
because small number statistics prohibit us making any firm statement on 
the radial variations of GC colors for the two fainter galaxies
after the correction for the NGC 1399 GC contamination. 
In order to investigate the radial variation of color bimodality,
we set up a series of radial bins containing approximately equal number ($\sim$\,300) of GCs 
and perform the GMM analysis for each bin.
We focus particularly on three main parameters of the GMM analysis 
that characterize the bimodal distributions; 
the number ratio between blue and red GCs, the mean colors of the groups, and their color dispersions. 

Figure~\ref{fig15_N1399r} shows    					
the variations of color bimodality properties as a function of galactocentric radius.
There is an obvious radial trend for the mean colors of both blue and red GCs
getting bluer with increasing radius in all colors.
The radial color gradient of red GCs is steeper than that of blue GCs.   
The red GC fraction in each color rapidly decreases 
with increasing galactocentric radius out to $r$ $\simeq$ $4\arcmin$, 
and it remains fairly constant beyond the point.
There is a weak radial trend in GC color dispersions,        
in that the ratio between dispersions of blue and red GCs 
increases with radius.

\section{SUMMARY AND CONCLUSIONS}

We have performed wide-field $UBVI$ photometry of GCs 
in the central region of the Fornax cluster of galaxies 
with the Mosaic II CCD imager on the 4 m Blanco telescope at CTIO. 
This is one of the widest and deepest $U$-band studies on extragalactic GC systems. 
The reduction was carried out with the DAOPHOT II/ALLFRAME package. 
Using two-color diagrams and the magnitude cut of $M_{V}\le-11$, 
a total of 2037 GC candidates were selected among 12,134 objects detected in all four bands. 
Our estimate of the contamination by background galaxies and foreground stars in our sample 
is in the range of 9\,\%\,--16\,\%.
The sample is $U$-band limited, and the completeness of our photometry is 50\,\% at $U_{0}=24.4$   
according to our artificial star tests. 
We provide the $UBVI$ photometric catalog of the GC candidates online. 
For the GC systems of NGC 1399, NGC 1404, and NGC 1387, 
we have investigated the spatial distributions, CMDs, and color distributions.  
For NGC 1399 GCs, the radial variations in the properties of bimodality were examined as well.  
The main results are summarized as follows.

\begin{enumerate}

\item 
We find an asymmetric distribution of GC candidates around NGC 1399.
The overabundance of GCs at the southeast side of the inner region 
and toward the directions of NGC 1404 and NGC 1387 in the outer region 
suggests that there were recent interactions of NGC 1399 with the galaxies
and that NGC 1399 is not yet dynamically relaxed.

\item 
We specify the radial extent of each GC system 
at the radius where the surface density profile begins to depart from the de Vaucouleurs fit. 
The limiting radii are $\sim$\,3\farcm4 for NGC 1404 and $\sim$\,3\farcm7 for NGC 1387. 
The GC system of NGC 1399 is spatially extended beyond the field of view of our observation.

\item  
We show that the GC systems in the three galaxies exhibit bimodal color distributions in all the colors considered in this study. 
For the brightest GCs in NGC 1399, the bimodality becomes weak or disappear depending on colors.

\item 
We find with GMM 
that the morphology of the color histograms changes significantly, depending on the colors used. 
The color dispersion of red GCs is larger than that of blue GCs in $U$-band colors  
by more than a factor of two, 
while the red and blue GCs show similar dispersions in $(B-V)_{0}$ and $(V-I)_{0}$. 
The number ratios between blue and red GCs vary depending on colors used as well.
When corrected for the contribution of NGC 1399 GCs, 
the red GC fractions of NGC 1404 and NGC 1387
remain constant regardless of the colors used. 

\item  
We suggest a possible scenario that could explain the presence of two discrete GC populations in NGC 1404 and NGC 1387:
(1) the central, red GCs have been preferentially left through the early interactions of the galaxies with NGC 1399 and
(2) later blue GCs have been accreted from outer region of NGC 1399.

\item   
We confirmed that the mean colors of blue and red GCs in NGC 1399
show a radial trend of becoming bluer with increasing galactocentric radius. 
The gradient of red GCs is steeper than that of blue GCs.

\end{enumerate}

\acknowledgments

S.-J.Y. acknowledges support 
from International Exchange Program for University Researchers (NRF-2011-013-C00031) 
and from Mid-career Research Program (No. 2012R1A2A2A01043870) 
through the National Research Foundation (NRF) of Korea grant funded 
by the Ministry of Education, Science and Technology (MEST), 
and support by the NRF of Korea to the Center for Galaxy Evolution Research (No. 2012-8-1743) 
and by the Korea Astronomy and Space Science Institute Research Fund 2011 and 2012. 
This work is partially supported by the KASI-Yonsei Joint
Research Program (2011--2012) for the Frontiers of Astronomy and
Space Science funded by the Korea Astronomy and Space Science Institute. 
This material is based upon work supported by AURA 
through the NSF under AURA Cooperative Agreement AST 0132798, as amended. 
\acknowledgments

\end{document}